\documentclass[twocolumn]{aastex62}
\usepackage{latexsym}
\usepackage{graphicx}
\usepackage{color}
\usepackage{amsmath}

\usepackage{hyperref}
\usepackage{bm}
\usepackage{acronym}
\usepackage{url}
\usepackage{textcomp}
\usepackage{xspace}
\usepackage{acronym}
\usepackage{multirow}

\newcommand{\macro}[1]{{#1}\xspace}

\newcommand{\NPREVPULSARS}{\macro{271}}  
\newcommand{\NPULSARS}{\macro{221}}
\newcommand{\NHIGHVALUE}{\macro{34}}  
\newcommand{\NBELOWTENHZ}{\macro{three}}  
\newcommand{\NFULLEMOVERLAP}{\macro{167}}  

\newcommand{\BESTPREVHPSR}{\macro{PSR\,J1918\textminus0642}}  
\newcommand{\BESTPREVHVAL}{\macro{\scinum{1.6}{-26}}}         
\newcommand{\BESTPREVELLPSR}{\macro{PSR\,J0636+5129}}         
\newcommand{\BESTPREVELLVAL}{\macro{\scinum{1.3}{-8}}}        
\newcommand{\NBELOWSPINDOWNPREV}{\macro{eight}}               
\newcommand{\CRABSDRATIOPREV}{\macro{20}}                     
\newcommand{\VELASDRATIOPREV}{\macro{9}}                     
\newcommand{\PULSAROVERLAP}{\macro{167}}                    
\newcommand{\PULSARNEW}{\macro{55}}                         

\newcommand{\OONESTARTDATE}{\macro{2015 September 11}}
\newcommand{\OONESTARTTIMELHO}{\macro{01:25:03~UTC}}
\newcommand{\OONESTARTTIMELLO}{\macro{18:29:03~UTC}}
\newcommand{\OONEENDDATE}{\macro{2016 January 19}}
\newcommand{\OONEENDTIME}{\macro{17:07:59~UTC}}
\newcommand{\OONEOBSTIMELHO}{\macro{79\,d}}
\newcommand{\OONEDUTYFACTORLHO}{\macro{60\%}}
\newcommand{\OONEOBSTIMELLO}{\macro{66\,d}}
\newcommand{\OONEDUTYFACTORLLO}{\macro{51\%}}
\newcommand{\OTWOSTARTDATE}{\macro{2016 November 30}}
\newcommand{\OTWOSTARTTIME}{\macro{16:00:00~UTC}}
\newcommand{\OTWOENDDATE}{\macro{2017 August 25}}
\newcommand{\OTWOENDTIME}{\macro{22:00:00~UTC}}

\newcommand{\OONECALAMPLHO}{\macro{5\%}}
\newcommand{\OONECALPHASELHO}{\macro{\ensuremath{3\degr}}}
\newcommand{\OONECALAMPLLO}{\macro{10\%}}
\newcommand{\OONECALPHASELLO}{\macro{\ensuremath{4\degr}}}
\newcommand{\OTWOCALAMPLHO}{\macro{3\%}}
\newcommand{\OTWOCALPHASELHO}{\macro{\ensuremath{3\degr}}}
\newcommand{\OTWOCALAMPLLO}{\macro{8\%}}
\newcommand{\OTWOCALPHASELLO}{\macro{\ensuremath{4\degr}}}

\newcommand{\HZEROSENS}{\macro{10.4}}  
\newcommand{\CTTSENS}{\macro{5.0}}     
\newcommand{\CTOSENS}{\macro{19.9}}    

\newcommand{\CRABHZEROBAYES}{\macro{\scinum{1.9}{-26}}}     
\newcommand{\CRABQTTBAYES}{\macro{\scinum{7.7}{32}\,kg\,m$^2$}}  
\newcommand{\CRABELLIPTICITYBAYES}{\macro{\scinum{1.0}{-5}}} 
\newcommand{\CRABHZEROFSTAT}{\macro{\scinum{2.2}{-26}}}     
\newcommand{\CRABHZEROFIVENVEC}{\macro{\scinum{2.9}{-26}}}  
\newcommand{\CRABSDBAYES}{\macro{0.013}}                    
\newcommand{\CRABPOWERBAYES}{\macro{0.017\%}}               
\newcommand{\VELAHZEROBAYES}{\macro{\scinum{1.4}{-25}}}     
\newcommand{\VELAQTTBAYES}{\macro{\scinum{5.9}{33}\,kg\,m$^2$}}  
\newcommand{\VELAELLIPTICITYBAYES}{\macro{\scinum{7.6}{-5}}} 
\newcommand{\VELAHZEROFSTAT}{\macro{\scinum{2.6}{-25}}}     
\newcommand{\VELAHZEROFIVENVEC}{\macro{\scinum{2.3}{-25}}}  
\newcommand{\VELASDBAYES}{\macro{0.042}}                    
\newcommand{\VELAPOWERBAYES}{\macro{0.18\%}}                

\newcommand{\NBELOWSPINDOWN}{\macro{20}}         
\newcommand{\NSPINDOWNONETOTEN}{\macro{53}}      
\newcommand{\NSPINDOWNBELOWTENMSP}{\macro{41}}   

\newcommand{\SMALLESTHZERO}{\macro{\scinum{8.9}{-27}}}      
\newcommand{\SMALLESTHZEROPSR}{\macro{J1623\textminus2631}} 
\newcommand{\SMALLESTHZERODIST}{\macro{1.8\,kpc}}           
\newcommand{\SMALLESTHZEROFREQ}{\macro{90.3\,Hz}}           

\newcommand{\SMALLESTQTT}{\macro{\scinum{4.5}{29}}}        
\newcommand{\SMALLESTELLIPTICITY}{\macro{\scinum{5.8}{-9}}} 
\newcommand{\SMALLESTQTTRATIO}{\macro{3.4}}                
\newcommand{\SMALLESTQTTPSR}{\macro{J0636+5129}}           
\newcommand{\SMALLESTQTTDIST}{\macro{0.21\,kpc}}           
\newcommand{\SMALLESTQTTFREQ}{\macro{348.6\,Hz}}         

\newcommand{\SMALLESTCTO}{\macro{\scinum{1.3}{-26}}}          
\newcommand{\SMALLESTCTOPSR}{\macro{J1744\textminus7619}}        
\newcommand{\SMALLESTCTOFREQ}{\macro{213.3\,Hz}}         

\newcommand{\SMALLESTMSPSDRAT}{\macro{1.3}}      
\newcommand{\SMALLESTMSPSDRATPSR}{\macro{J0711\textminus6830}} 
\newcommand{\SMALLESTMSPSDRATDIST}{\macro{0.11\,kpc}} 
\newcommand{\SMALLESTMSPSDRATFREQ}{\macro{182.1\,Hz}} 
\newcommand{\SMALLESTMSPSDRATHZERO}{\macro{\scinum{1.5}{-26}}} 
\newcommand{\SMALLESTMSPSDRATQTT}{\macro{\scinum{9.3}{29}\,kg\,m$^2$}} 
\newcommand{\SMALLESTMSPSDRATELLIPTICITY}{\macro{\scinum{1.2}{-8}}} 

\newcommand{\gw}{gravitational-wave\xspace}

\newcommand{\scinum}[2]{\ensuremath{#1\!\times\!10^{#2}}}

\acrodef{MSP}[MSP]{millisecond pulsar}

\begin{document}

\reportnum{LIGO-P1800344}

\title{Searches for Gravitational Waves from Known Pulsars at Two Harmonics in 2015--2017 LIGO Data}


\author{B.~P.~Abbott}
\affiliation{LIGO, California Institute of Technology, Pasadena, CA 91125, USA}
\author{R.~Abbott}
\affiliation{LIGO, California Institute of Technology, Pasadena, CA 91125, USA}
\author{T.~D.~Abbott}
\affiliation{Louisiana State University, Baton Rouge, LA 70803, USA}
\author{S.~Abraham}
\affiliation{Inter-University Centre for Astronomy and Astrophysics, Pune 411007, India}
\author{F.~Acernese}
\affiliation{Universit\`a di Salerno, Fisciano, I-84084 Salerno, Italy}
\affiliation{INFN, Sezione di Napoli, Complesso Universitario di Monte S.Angelo, I-80126 Napoli, Italy}
\author{K.~Ackley}
\affiliation{OzGrav, School of Physics \& Astronomy, Monash University, Clayton 3800, Victoria, Australia}
\author{C.~Adams}
\affiliation{LIGO Livingston Observatory, Livingston, LA 70754, USA}
\author{R.~X.~Adhikari}
\affiliation{LIGO, California Institute of Technology, Pasadena, CA 91125, USA}
\author{V.~B.~Adya}
\affiliation{Max Planck Institute for Gravitational Physics (Albert Einstein Institute), D-30167 Hannover, Germany}
\affiliation{Leibniz Universit\"at Hannover, D-30167 Hannover, Germany}
\author{C.~Affeldt}
\affiliation{Max Planck Institute for Gravitational Physics (Albert Einstein Institute), D-30167 Hannover, Germany}
\affiliation{Leibniz Universit\"at Hannover, D-30167 Hannover, Germany}
% ...

\collaboration{The LIGO Scientific Collaboration and the Virgo Collaboration}


\author{Z.~Arzoumanian}
\affiliation{X-Ray Astrophysics Laboratory, NASA Goddard Space Flight Center, Greenbelt, MD 20771, USA}
\author{S.~Bogdanov}
\affiliation{Columbia Astrophysics Laboratory, Columbia University, 550 West 120th Street, New York, NY, 10027, USA}
\author{I.~Cognard}
\affiliation{Laboratoire de Physique et Chimie de l'Environnement et de l'Espace -- Universit\'{e} d'Orl\'{e}ans / CNRS, F-45071 Orl\'{e}ans Cedex 02, France}
\affiliation{Station de Radioastronomie de Nan\c{c}ay, Observatoire de Paris, CNRS/INSU, F-18330 Nan\c{c}ay, France}
\author{A.~Corongiu}
\affiliation{INAF--Osservatorio Astronomico di Cagliari, via della Scienza 5, 09047 Selargius, Italy}
\author{T.~Enoto}
\affiliation{Hakubi Center for Advanced Research and Department of Astronomy, Kyoto University, Kyoto 606-8302, Japan}
\author{P.~Freire}
\affiliation{Max-Planck-Institut f\"{u}r Radioastronomie, Auf dem H\"{u}gel 69, D-53121 Bonn, Germany}
\author{K.~C.~Gendreau}
\affiliation{X-Ray Astrophysics Laboratory, NASA Goddard Space Flight Center, Greenbelt, MD 20771, USA}
\author{L.~Guillemot}
\affiliation{Laboratoire de Physique et Chimie de l'Environnement et de l'Espace -- Universit\'{e} d'Orl\'{e}ans / CNRS, F-45071 Orl\'{e}ans Cedex 02, France}
\affiliation{Station de Radioastronomie de Nan\c{c}ay, Observatoire de Paris, CNRS/INSU, F-18330 Nan\c{c}ay, France}
\author{A.~K.~Harding}
\affiliation{Astrophysics Science Division, NASA Goddard Space Flight Center, Greenbelt, MD 20771, USA}
\author{F.~Jankowski}
\affiliation{Jodrell Bank Centre for Astrophysics, School of Physics and Astronomy, University of Manchester, Manchester, M13 9PL, UK}
\author{M.~J.~Keith}
\affiliation{Jodrell Bank Centre for Astrophysics, School of Physics and Astronomy, University of Manchester, Manchester, M13 9PL, UK}
\author{M.~Kerr}
\affiliation{Space Science Division, Naval Research Laboratory, Washington, DC 20375-5352, USA}
\author{A.~Lyne}
\affiliation{Jodrell Bank Centre for Astrophysics, School of Physics and Astronomy, University of Manchester, Manchester, M13 9PL, UK}
\author{J.~Palfreyman}
\affiliation{Department of Physical Sciences, University of Tasmania, Private Bag 37, Hobart, Tasmania 7001, Australia}
\author{A.~Possenti}
\affiliation{INAF--Osservatorio Astronomico di Cagliari, via della Scienza 5, 09047 Selargius, Italy}
\affiliation{Universit\`{a} di Cagliari, Dipartimento di Fisica, I-09042, Monserrato, Italy}
\author{A.~Ridolfi}
\affiliation{Max-Planck-Institut f\"{u}r Radioastronomie, Auf dem H\"{u}gel 69, D-53121 Bonn, Germany}
\author{B.~Stappers}
\affiliation{Jodrell Bank Centre for Astrophysics, School of Physics and Astronomy, University of Manchester, Manchester, M13 9PL, UK}
\author{G.~Theureau}
\affiliation{Laboratoire de Physique et Chimie de l'Environnement et de l'Espace -- Universit\'{e} d'Orl\'{e}ans / CNRS, F-45071 Orl\'{e}ans Cedex 02, France}
\affiliation{Station de Radioastronomie de Nan\c{c}ay, Observatoire de Paris, CNRS/INSU, F-18330 Nan\c{c}ay, France}
\affiliation{LUTH, Observatoire de Paris, PSL Research University, CNRS, Universit\'{e} Paris Diderot, Sorbonne Paris Cit\'{e}, F-92195 Meudon, France}
\author{P.~Weltervrede}
\affiliation{Jodrell Bank Centre for Astrophysics, School of Physics and Astronomy, University of Manchester, Manchester, M13 9PL, UK}

\date{\today}

\begin{abstract}
We present a search for gravitational waves from \NPULSARS pulsars with rotation frequencies
$\gtrsim 10$\,Hz. We use advanced LIGO data from its first and second observing runs spanning
2015--2017, which provides the highest-sensitivity \gw data so far obtained. In this search we target
emission from both the $l=m=2$ mass quadrupole mode, with a frequency at twice that of the pulsar's
rotation, and from the $l=2$, $m=1$ mode, with a frequency at the pulsar rotation frequency. The
search finds no evidence for gravitational-wave emission from any pulsar at either frequency. For
the $l=m=2$ mode search, we provide updated upper limits on the gravitational-wave amplitude, mass
quadrupole moment, and fiducial ellipticity for \PULSAROVERLAP pulsars, and the first such limits
for a further \PULSARNEW. For \NBELOWSPINDOWN young pulsars these results give limits that are below
those inferred from the pulsars' spin-down. For the Crab and Vela pulsars our results constrain
gravitational-wave emission to account for less than \CRABPOWERBAYES and \VELAPOWERBAYES of the
spin-down luminosity, respectively. For the recycled millisecond pulsar \SMALLESTMSPSDRATPSR our
limits are only a factor of \SMALLESTMSPSDRAT above the spin-down limit, assuming the canonical
value of $10^{38}\,\text{kg}\,\text{m}^2$ for the star's moment of inertia, and imply a
gravitational-wave-derived upper limit on the star's ellipticity of \SMALLESTMSPSDRATELLIPTICITY. We
also place new limits on the emission amplitude at the rotation frequency of the pulsars.
\end{abstract}


\section{Introduction}

There have been several previous searches for persistent (or continuous) quasi-monochromatic
gravitational waves emitted by a selection of known pulsars using data from the LIGO, Virgo, and
GEO600 \gw detectors
\citep{2004PhRvD..69h2004A,2005PhRvL..94r1103A,2007PhRvD..76d2001A,2008ApJ...683L..45A,2010ApJ...713..671A,2011ApJ...737...93A,2014ApJ...785..119A,2017ApJ...839...12A}.
In the majority of these, the signals that have been searched for are those that would be expected
from stars with a nonzero $l=m=2$ mass quadrupole moment $Q_{22}$ and with polarization content
consistent with the expectations of general relativity \citep[see,
e.g.,][]{1979PhRvD..20..351Z,1996AA...312..675B,1998PhRvD..58f3001J}. Such signals would be produced
at twice the stellar rotation frequencies, and searches have generally assumed that the rotation
frequency derived from electromagnetic observations of the pulsars is phase locked to the star's
rotation and thus the \gw signal. Some searches have been performed where the assumption of the
phase locking to the observed electromagnetic signal has been slightly relaxed, allowing the signal
to be potentially offset over a small range of frequencies ($\sim 10-100$\,mHz) and first frequency
derivatives \citep{2008ApJ...683L..45A,2015PhRvD..91b2004A,2017PhRvD..96l2006A}. A search including
the prospect of the signal's polarization content deviating from the purely tensorial modes
predicted by general relativity has also been performed in \citet{2018PhRvL.120c1104A}. None of
these searches have detected a \gw signal from any of the pulsars that were targeted. Thus,
stringent upper limits of the \gw amplitude, mass quadrupole moment, and ellipticity have been set.

Emission of gravitational waves at a pulsar's rotation frequency from the $l=2$, $m=1$ harmonic
mode, in addition to emission at twice the rotation frequency from the $l=m=2$ mode, has long been
theorized \citep{1979PhRvD..20..351Z, 1980PhRvD..21..891Z,2002MNRAS.331..203J}. The fiducial
emission mechanism would be from a biaxial, or triaxial star, undergoing free precession. In the
case of a precessing biaxial star, or a precessing triaxial star with a small ``wobble angle,'' the
electromagnetic pulsar emission frequency would be modulated slightly, with the \gw emission being
emitted at frequencies close to once and twice the time-averaged rotation frequency. There is only
weak observational evidence for any pulsar showing precession \citep[see the discussions in,
e.g.,][and references therein]{2012MNRAS.420.2325J,2013ApJ...763...72D}, and free precession
would be quickly damped, but as shown in \citet{2010MNRAS.402.2503J} the existence of a superfluid
interior gives rise to the possibility for \gw emission at the rotation frequency even for a
nonprecessing star. A search for emission at both once and twice the rotation frequency for 43
pulsars using data from LIGO's fifth science run has been performed in \citet{2015MNRAS.453.4399P}.
That analysis saw no evidence for signals at the rotation frequency and was consistent with the
search conducted for signals purely from the $l=m=2$ mode \citep{2010ApJ...713..671A}.

The searches implemented in this work are specifically designed for the case where the signal's
phase evolution is very well known over the course of full \gw detector observing runs. Therefore,
here we will only focus on the assumption that emission occurs at precisely once and twice the
observed rotation frequency, as given by the model in \citet{2010MNRAS.402.2503J}, so we do not
account for the possibility of any of the sources undergoing free precession.

Previous searches, combining the results given in \citet{2014ApJ...785..119A} and
\citet{2017ApJ...839...12A}, have included a total of \NPREVPULSARS pulsars. The most stringent
upper limit on \gw amplitude from the $l=m=2$ mode was set for \BESTPREVHPSR at \BESTPREVHVAL, and
the most stringent upper limit on the fiducial ellipticity (see Appendix~\ref{ap:definitions},
Equations~(\ref{eq:ell}) and (\ref{eq:fidell})) was set for \BESTPREVELLPSR at \BESTPREVELLVAL
\citep{2017ApJ...839...12A}. However, for these particular pulsars, both of which are \acp{MSP}, the
\gw amplitude limits are above the fiducial spin-down limit (see Appendix~\ref{ap:definitions} and
Equation~(\ref{eq:h0sd})). In the search described in \citet{2017ApJ...839...12A}, there
were \NBELOWSPINDOWNPREV pulsars for which their observed gravitational-wave limits were
below the fiducial spin-down limits, with the upper limits on emission from the Crab pulsar
(PSR\,J0534+2200) and Vela pulsar (PSR\,J0835\textminus4510) being factors of more than
\CRABSDRATIOPREV and \VELASDRATIOPREV below their respective spin-down limits.\footnote{In
previous work we have often referred to observed gravitational-wave limits ``surpassing,'' or
``beating,'' the spin-down limits, which just means to say that the limits are lower than the
equivalent spin-down limits.}

Concurrently with this work, a search has been performed for 33 pulsars using advanced LIGO data
from the second observing run in which the assumption of phase locking between the
electromagnetically observed signal and \gw signal is relaxed by allowing the signal model to vary
freely over a narrow band of frequencies and frequency derivatives \citep{O2Narrowband}.
Even with the slight sensitivity decrease compared to the analysis presented here, due to the wider
parameter space, that analysis gives limits that are below the spin-down limit for 13 of the pulsars.

\subsection{Signal model}

Using the formalism shown in \citet{2015MNRAS.453...53J} and \citet{2015MNRAS.453.4399P} the \gw
waveform from the $l=2$, $m=1$ harmonic mode can be written as
\begin{widetext}
\begin{equation}\label{eq:l2m1}
h_{21}(t) = -\frac{C_{21}}{2}\left[F^D_+(\alpha, \delta, \psi; t)\sin\iota\cos\iota\cos{\left(\Phi(t)+\Phi_{21}^C\right)} + F^D_{\times}(\alpha, \delta, \psi; t)\sin\iota \sin{\left(\Phi(t)+\Phi_{21}^C\right)}\right],
\end{equation}
and that from the $l=m=2$ mode can be written as
\begin{equation}\label{eq:l2m2}
h_{22}(t) = -C_{22}\left[F^D_+(\alpha, \delta, \psi; t)\left(1+\cos{}^2\iota\right)\cos{\left(2\Phi(t)+\Phi_{22}^C\right)} + 2F^D_{\times}(\alpha, \delta, \psi; t)\cos\iota \sin{\left(2\Phi(t)+\Phi_{22}^C\right)}\right].
\end{equation}
\end{widetext}
Here $C_{21}$ and $C_{22}$ represent the amplitudes of the components, $\Phi_{21}^C$ and
$\Phi_{22}^C$ represent initial phases at a particular epoch, $\Phi(t)$ is the rotational phase of
the source, and $\iota$ is the inclination of the source's rotation axis with respect to the
line of sight.\footnote{For precessing stars the phase evolution $\Phi(t)$ in
Equations~(\ref{eq:l2m1}) and (\ref{eq:l2m2}) will not necessarily be given by the rotational phase,
but can differ by the precession frequency.} The detected amplitude is modulated by the detector
response functions for the two polarizations of the signal (`+' and `$\times$'), $F^D_+(\alpha,
\delta, \psi; t)$ and $F^D_{\times}(\alpha, \delta, \psi; t)$, which depend on the location and
orientation of detector $D$, the location of the source on the sky, defined by the R.A.\
$\alpha$ and decl.\ $\delta$, and the polarization angle of the source $\psi$.

As shown in \citet{2015MNRAS.453...53J}, the waveforms given in Equations~(\ref{eq:l2m1}) and
(\ref{eq:l2m2}) describe a generic signal, but the amplitudes ($C_{21}$ and $C_{22}$) and phases
($\Phi_{21}^C$ and $\Phi_{22}^C$) can be related to intrinsic physical parameters describing a
variety of source models, e.g., a triaxial star spinning about a principal axis
\citep{2004PhRvD..69h2004A}, a biaxial precessing star \citep{2002MNRAS.331..203J}, or a triaxial
star not spinning about a principal axis \citep{2010MNRAS.402.2503J}. In the standard case adopted
for previous \gw searches of a triaxial star spinning about a principal axis, there is only emission
at twice the rotation frequency from the $l=m=2$ mode, so only Equation~(\ref{eq:l2m2}) is non-zero.
In this case the $C_{22}$ amplitude can be simply related to the standard \gw strain amplitude $h_0$
via $h_0 = 2C_{22}$.\footnote{To maintain the sign convention between Equation~(\ref{eq:l2m2}) and
the equivalent equation in, e.g., \citet{1998PhRvD..58f3001J}, the transform between $h_0$ and
$C_{22}$ should more strictly be $h_0 = -2C_{22}$.} We can simply define the phase $\Phi_{22}^C$ as
relating to the initial rotational phase $\phi_0$ via $\Phi_{22}^C = 2\phi_0$, noting that $\phi_0$
actually incorporates the sum of two phase parameters (an initial \gw phase and another phase
offset) that are entirely degenerate and therefore not separately distinguishable
\citep{2015MNRAS.453...53J}.

Despite Equations~(\ref{eq:l2m1}) and (\ref{eq:l2m2}) not providing the intrinsic parameters of the
source, they do break strong degeneracies between them, which are otherwise impossible to
disentangle \citep[see][showing this for the case of a triaxial source not rotating about a
principal axis]{2015MNRAS.453.4399P}.

In this work we adopt two analyses. The first assumes the standard picture of a triaxial star
rotating around a principal axis from which we can simply relate the waveform amplitude $C_{22}$ to
the \gw amplitude. In this case we can then compare this to the standard spin-down limit and can
calculate each source's mass quadrupole $Q_{22}$ and fiducial ellipticity upper limits (see
Appendix~\ref{ap:definitions} for definitions of these standard quantities.) The second assumes the
model of a triaxial star not spinning about a principal axis, for which there could be emission at
both once or twice the rotation frequency. In this case we do not attempt to relate the signal
amplitudes to any physical parameter of the source.

\subsection{Signal strength}\label{sec:strength}

For the $l=m=2$ quadrupole mode the strength of the emission is defined by the size of the mass
quadrupole moment $Q_{22}$ (see Equations~(\ref{eq:h0}) and (\ref{eq:q22})), which is proportional
to the ellipticity of the star and to the star's moment of inertia, and will therefore depend upon
the star's mass and also upon the equation of state of neutron star matter \citep[see,
e.g.,][]{2000MNRAS.319..902U,2005PhRvL..95u1101O,2013PhRvD..88d4004J}. This ellipticity could be
provided by some physical distortion of the star's crust or irregularities in the density profile of
the star. For our purposes the mechanism providing the distortion must be sustained over long
periods, e.g., the crust must be strong enough for any (submillimetre high) mountain to be
maintained \citep[see][for discussions of the maximum sustainable ellipticities for various neutron
star equations of state]{2005PhRvL..95u1101O,2013PhRvD..88d4004J}, or there must be a persistent
strong internal magnetic field \citep[e.g.,][]{1996AA...312..675B,2002PhRvD..66h4025C}.
\citet{2013PhRvD..88d4004J} suggest that, assuming a standard set of neutron star
equations of state, maximum fiducial ellipticities of a few $\times 10^{-6}$ could be sustained.
Constraints on the neutron star equation of state are now starting to be probed using \gw
observations from the binary neutron star coalescence observed as GW170817
\citep{2017PhRvL.119p1101A,2018PhRvL.121p1101A}. These constraints suggest that softer
equations of state are favored over stiffer ones, which would imply smaller maximum crustal
quadrupoles. An additional caveat to this is that the maximum crustal deformation is also dependent
on the star's mass, and less massive stars would allow larger deformations
\citep{2010PhRvD..81j3001H,2013PhRvD..88d4004J}, so there is still a wide range of uncertainty.
Recent work on the strength of neutron star crusts consisting of nuclear pasta suggests that these
could have larger breaking strains and thus support larger ellipticities
\citep{2018PhRvL.121m2701C}.

It has recently been suggested by \citet{2018ApJ...863L..40W} that the distribution of \acp{MSP} in
the period--period derivative plane provides some observational evidence that they may all have a
limiting minimum ellipticity of $\sim 10^{-9}$. This could be due to some common process that
takes place during the recycling accretion stage that spins the pulsar up to millisecond periods.
For example, there could be external magnetic field burial \citep[see,
e.g.,][]{2001PASA...18..421M,2004MNRAS.351..569P} for which the size of the buried field is roughly
the same across all stars, or similar levels of spin-up leading to crust breaking
\citep[e.g.,][]{2018arXiv180404952F}. If this is true, it provides a compelling reason to look for
emission from these objects.

For the model emitting at both $l=2$, $m=1,2$ modes, and assuming no precession, the signal
amplitudes are related to combinations of moment-of-inertia asymmetries and orientation angles
between the crust and core of the star \citep{2010MNRAS.402.2503J}. These are related in a complex
way to the $C_{21}$ and $C_{22}$ amplitudes given in Equations~(\ref{eq:l2m1}) and (\ref{eq:l2m2})
\citep[see][]{2015MNRAS.453...53J}. In general, if the $Q_{21}$ and $Q_{22}$ mass moments are equal,
then the \gw strain from the $l=2$, $m=1$ mode would be roughly four times smaller owing to the fact
that it is related to the square of the frequency and that mode is at half the frequency of the
$l=m=2$ mode. However, we do not have good estimates of what the actual relative mass moments might
be.

Note that one can in principle also obtain limits on a neutron star's deformation if one interprets
some features of its timing properties as due to free precession.  In this case, the limits involve
a combination of the differences between the three principal moments of inertia, together with an
angular parameter (``wobble angle'') giving the amplitude of the precession.  This can be done either
for stars that show some periodic structure in their timing properties \citep[see,
e.g.,][]{2006MNRAS.365..653A, 2017MNRAS.467..164A}, or by assuming that some component of pulsar
timing noise is due to precession \citep{1993ASPC...36...43C}. Note, however, that it is by no means
clear whether pulsar timing really does provide evidence for free precession
\citep{2017PhRvL.118z1101J, 2019MNRAS.tmp..628S}.

\subsection{Search methods}\label{sec:methods}

As with the previous searches for gravitational waves from known pulsars described in
\citet{2014ApJ...785..119A} and \citet{2017ApJ...839...12A}, we make use of three semi-independent
search methods. We will not describe these methods in detail here, but refer the reader to
\citet{2014ApJ...785..119A} for more information. Briefly, the three methods are as follows: a search using
narrowband time-domain data to perform Bayesian parameter estimation for the unknown signal
parameters, and marginal likelihood evaluation, for each pulsar
\citep{Dupuis:2005,2017arXiv170508978P}; a search using the same narrow-banded time series, but
Fourier-transformed into the frequency domain, to calculate the $\mathcal{F}$-statistic
\citep{1998PhRvD..58f3001J} \citep[or equivalent $\mathcal{G}$-statistic for constrained
orientations;][]{Jaranowski:2010}, with a frequentist-based amplitude upper limit estimation
procedure \citep{1998PhRvD..57.3873F}; and a search in the frequency domain that makes use of
splitting of any astrophysical signal into five frequency harmonics through the sidereal amplitude
modulation given by the detector responses \citep{2010CQGra..27s4016A,Astone:2012}. The
narrowband time-domain data are produced by heterodyning the raw detector strain data using the
expected signal's phase evolution \citep{Dupuis:2005}. It is then low-pass-filtered with a knee
frequency of 0.25\,Hz and downsampled, via averaging, creating a complex time series with one
sample per minute, i.e., a bandwidth of $1/60$\,Hz centered about the expected signal frequency that
is now at 0\,Hz. We call these approaches the {\it Bayesian}, $\mathcal{F}$-/$\mathcal{G}$-{\it
statistic}, and $5n$-{\it vector} methods, respectively. The first of these methods has been applied
to all the pulsars in the sample (see Section~\ref{sec:pulsars}), and again following
\citet{2014ApJ...785..119A} and \citet{2017ApJ...839...12A} at least two of the above methods have been applied
to a selection of \NHIGHVALUE high-value targets for which the observed limit is lower than,
or closely approaches, the spin-down limit. The results of the $5n$-{\it vector} analysis only use
data from the LIGO O2 run (see Section~\ref{sec:gwdata}).

All these methods have been adapted to deal with the potential for signals at both once and twice
the rotation frequency. For the {\it Bayesian} method, when searching for such a signal the
narrowband time series from both frequencies are included in a coherent manner, with common
polarization angles $\psi$ and orientations $\iota$. For the $5n$-{\it vector} and
$\mathcal{F}$-/$\mathcal{G}$-{\it statistic} methods a simpler approach is taken, and signals at the
two frequencies are searched for independently. The $\mathcal{F}$/$\mathcal{G}$-{\it statistic}
approach for such a signal is described in more detail in \citet{2014CQGra..31j5011B}. As a
consequence, given that $C_{21}=0$ (see Equation~(\ref{eq:l2m1})) corresponds to the case of a
triaxial star rotating around one of its principal axes of inertia, results for the amplitude
$C_{22}$ (Equation~(\ref{eq:l2m2})) from the $5n$-{\it vector} method are not given, as they are
equivalent to those for the standard amplitude $h_0$.

In the case of a pulsar being observed to glitch during the run (see Section~\ref{sec:pulsars}) the
methods take different approaches. For the {\it Bayesian} method it is assumed that any glitch may
produce an unknown offset between the electromagnetically observed rotational phase and the \gw
phase. Therefore, an additional phase offset is added to the signal model at the time of the glitch,
and this is included as a parameter to be estimated, while the \gw amplitude and orientation angles
of the source (inclination and polarization) are assumed to remain fixed over the glitch. This is
consistent with the analysis in \citet{2010ApJ...713..671A}, although it differs from the more recent
analyses in \citet{2014ApJ...785..119A} and \citet{2017ApJ...839...12A} in which each interglitch period was
treated semi-independently, i.e., independent phases and polarization angles were assumed for each
interglitch period, but two-dimensional marginalized posterior distributions on the \gw amplitude
and cosine of the inclination angle from data before a glitch were used as a prior on those
parameters when analyzing data after the glitch. For both the $\mathcal{F}$/$\mathcal{G}$-{\it
statistic} and $5n$-{\it vector} methods, as already done in
\citet{2014ApJ...785..119A} and \citet{2017ApJ...839...12A}, each interglitch period is analyzed independently,
i.e., no parameters are assumed to be coherent over the glitch, and the resulting statistics are
incoherently combined.

The prior probability distributions for the unknown signal parameters, as used for the {\it
Bayesian} and $5n$-{\it vector} methods, are described in Appendix~\ref{ap:priors}.

The $5n$-{\it vector} method uses a description of the \gw signal based on the concept of
polarization ellipse. The relation of the amplitude parameter $H_0$ used by the $5n$-{\it vector}
method with both the standard strain amplitude $h_0$ and the $C_{21}$ amplitude given in
Equation~(\ref{eq:l2m1}) is described in Appendix~\ref{ap:5vecampl}.

\section{Data}

In this section we briefly detail both the \gw data that have been used in the searches and the
electromagnetic ephemerides for the selection of pulsars that have been included.

\subsection{Gravitational-wave data}\label{sec:gwdata}

The data analyzed in this paper consist of those obtained by the two LIGO detectors (the LIGO
Hanford Observatory, commonly abbreviated to LHO or H1, and the LIGO Livingston Observatory,
abbreviated to LLO or L1) taken during their first \citep{2016PhRvL.116m1103A} and second observing
runs (O1 and O2, respectively) in their advanced detector configurations \citep{2015CQGra..32g4001L}.\footnote{The O1 and O2 datasets
are publicly available via the Gravitational Wave Open Science Center at
\url{https://www.gw-openscience.org/O1} and \url{https://www.gw-openscience.org/O2}, respectively \citep{2015JPhCS.610a2021V}.}

Data from O1 between \OONESTARTDATE (with start times of \OONESTARTTIMELHO and \OONESTARTTIMELLO for
LHO and LLO, respectively) and \OONEENDDATE at \OONEENDTIME have been used. The calibration of these
data and the frequency-dependent uncertainties on amplitude and phase over the run are described in
detail in \citet{2017PhRvD..96j2001C}. Over the course of the O1 run the calibration amplitude
uncertainty was no larger than \OONECALAMPLHO and \OONECALAMPLLO, and the phase uncertainty was no
larger than \OONECALPHASELHO and \OONECALPHASELLO, for LHO and LLO, respectively, over the frequency
range $\sim 10-2000$\,Hz \citep[these are derived from the 68\% confidence levels given in Figure~11
of][]{2017PhRvD..96j2001C}. All data flagged as in ``science mode,'' i.e., when the detectors were
operating in a stable state, and for which the calibration was behaving as expected, have been used.
This gave a total of \OONEOBSTIMELHO and \OONEOBSTIMELLO observing time for LHO and LLO,
respectively, equivalent to duty factors of \OONEDUTYFACTORLHO and \OONEDUTYFACTORLLO. 

Data from O2 between \OTWOSTARTDATE at \OTWOSTARTTIME and \OTWOENDDATE at \OTWOENDTIME, for both LHO
and LLO, have been used. An earlier version of the calibrated data for this observing run, as well
as the uncertainty budget associated with it, is again described in \citet{2017PhRvD..96j2001C}.
However, data with an updated calibration has been produced and used in this analysis, with this
having an improved uncertainty budget \citep{Kissel:2018}. Over the course of the O2 run the
calibration amplitude uncertainty was no larger than \OTWOCALAMPLHO and \OTWOCALAMPLLO and the
phase uncertainty was no larger than \OTWOCALPHASELHO and \OTWOCALPHASELLO for LHO and LLO,
respectively, over the frequency range of $\sim 10-2000$\,Hz. The data used in this analysis were
post-processed to remove spurious jitter noise that affected detector sensitivity across a broad
range of frequencies, particularly for data from LHO, and to remove some instrumental spectral lines
\citep{2018arXiv180905348D,2018arXiv180600532D}.

The Virgo gravitational-wave detector \citep{2015CQGra..32b4001A} was operating during the last 25
days of O2 \citep{2017PhRvL.119n1101A}; however, due to its higher noise levels as compared to the
LIGO detectors and the shorter observing time, Virgo data were not included in this analysis.

\subsection{Pulsars}\label{sec:pulsars}

For this analysis we have gathered ephemerides for \NPULSARS pulsars based on radio, X-ray, and
$\gamma$-ray observations. The observations have used the 42\,ft telescope and Lovell telescope at
Jodrell Bank (UK), the Mount Pleasant Observatory 26\,m telescope (Australia), the Parkes radio
telescope (Australia), the Nan\c{c}ay Decimetric Radio Telescope (France), the Molonglo Observatory
Synthesis Telescope (Australia), the Arecibo Observatory (Puerto Rico), the Fermi Large Area
Telescope, and the Neutron Star Interior Composition Explorer (NICER). As with the search in
\citet{2017ApJ...839...12A}, the criterion for our selection of pulsars was that they have rotation
frequencies greater than 10\,Hz, so that they are within the frequency band of greatest sensitivity
of the LIGO instruments, and for which the calibration is well characterized. There are in fact
\NBELOWTENHZ pulsars with rotation frequencies just below 10~Hz that we include (PSR J0117+5914, PSR
J1826\textminus1256, and PSR J2129+1210A); for two of these the spin-down limit was potentially
within reach using our data.

The ephemerides have been created using pulse time-of-arrival observations that mainly overlapped
with all, or some fraction of, the O1 and O2 observing periods (see Section~\ref{sec:gwdata}), so
the timing solutions should provide coherent phase models over and between the two runs. Of the
\NPULSARS, we have \NFULLEMOVERLAP for which the electromagnetic timings fully overlapped with the
full O1 and O2 runs. There are 12 pulsars for which there is no overlap between electromagnetic
observations and the O2 run. These include two pulsars, J1412+7922 (known as Calvera) and
J1849\textminus0001, for which we only have X-ray timing observations from after O2
\citep{2019arXiv190200144B}.\footnote{Subsequent to the search performed here, \citet{2019arXiv190200144B}
revised their initial timing model of J1849\textminus0001 so that it now overlaps partially with O2.
The revised model is consistent with the initial model used here, and thus the results presented
here remain valid.} For these we have made the reasonable assumption that timing models are
coherent for our analysis and that no timing irregularities, such as glitches, are present.

In all previous searches a total of \NPREVPULSARS pulsars had been searched for, with \PULSAROVERLAP
of these being timed for this search. For the other sources ephemerides were not available to us
for our current analysis. In particular, we do not have up-to-date ephemerides for many of the
pulsars in the globular clusters 47~Tucanae and Terzan~5, or the interesting young X-ray pulsar
J0537\textminus6910.

\subsubsection{Glitches}

During the course of the O2 period, five pulsars exhibited timing glitches. The Vela pulsar
(J0835\textminus4510) glitched on 2016 December 12 at 11:36~UTC
\citep{2016ATel.9847....1P,2018Natur.556..219P}, and the Crab pulsar (J0534+2200) showed a small
glitch on 2017 March 27 at around 22:04~UTC
\citep{2011MNRAS.414.1679E}.\footnote{\url{http://www.jb.man.ac.uk/pulsar/glitches.html}}
PSR~J1028\textminus5819 glitched some time around 2017 May 29, with a best-fit glitch time of 01:36
UTC. PSR~J1718\textminus3825 experienced a small glitch around 2017 July 2. PSR~J0205+6449
experienced four glitches over the period between the start of O1 and the end of O2, with glitch
epochs of 2015 November 19, 2016 July 1, 2016 October 19, and 2017 May 27. Two of these glitches
occurred in the period between O1 and O2, and as such any effect of the glitches on discrepancies
between the electromagnetic and \gw phase would not be independently distinguishable, meaning that
effectively only three glitches need to be accounted for.

\subsubsection{Timing noise}

Timing noise is low-frequency noise observed in the residuals of pulsar pulse arrival times
after subtracting a low-order Taylor expansion fit \citep[see, e.g.,][]{Hobbs_2006}. As shown in
\citet{1980ApJ...239..640C}, \citet{1994ApJ...422..671A} timing noise is strongly correlated with pulsar
period derivative, so ``young,'' or canonical, pulsars generally have far higher levels than
\acp{MSP}. If not accounted for in the timing model, the Crab pulsar's phase, for example, could
deviate by on the order of a cycle over the course of our observations, leading to decoherence of
the signal \citep[see][]{2004PhRvD..70d2002J, 2007PhRvD..76d2006P, 2015PhRvD..91f2009A}. In our \gw
searches we used phase models that incorporate the effects of timing noise when necessary. In some
cases this is achieved by using a phase model that includes high-order coefficients in the Taylor
expansion (including up to the twelfth frequency derivative in the case of the Crab pulsar) when
fitting the electromagnetic pulse arrival times. In others, where expansions in the phase do not
perform well, we have used the method of fitting multiple sinusoidal harmonics to the timing noise
in the arrival times, as described in
\citet{2004MNRAS.353.1311H} and implemented in the {\sc Fitwaves} algorithm in {\sc Tempo2}
\citep{2006MNRAS.369..655H}.

\subsubsection{Distances and period derivatives}

When calculating results of the searches in terms of the $Q_{22}$ mass quadrupole, fiducial
ellipticity, or spin-down limits (see Appendix~\ref{ap:definitions}), we require the distances to the
pulsars. For the majority of pulsars we use ``best-estimate'' distances given in the ATNF Pulsar
Catalog \citep{2005AJ....129.1993M}.\footnote{Version 1.59 of the catalog available at
\url{http://www.atnf.csiro.au/people/pulsar/psrcat/}.} In the majority of cases these are distances
based on the observed dispersion measure and calculated using the Galactic electron density
distribution model of \citet{2017ApJ...835...29Y}, although others are based on parallax
measurements, or inferred from associations with other objects or flux measurements. The distances
used for each pulsar, as well as the reference for the value used, are given in Tables~\ref{tab:highvalue}
and \ref{tab:results}.

The spin-down limits that we compare our results to (see Appendix~\ref{ap:definitions}) require a
value for the first period derivative $\dot{P}$, or equivalently frequency derivative $\dot{f}$, of
the pulsar. The observed spin-down does not necessarily reflect the intrinsic spin-down of the
pulsar, as it can be contaminated by the relative motion of the pulsar with respect to the observer.
This is particularly prevalent for \acp{MSP}, which have intrinsically small spin-downs that can be
strongly affected, particularly if they are in the core of a globular cluster where significant
intracluster accelerations can occur, or if they have a large transverse velocity with respect to
the solar system and/or are close (the ``Shklovskii effect''; \citealt{1970SvA....13..562S}.) The
spin-down can also be contaminated by the differential motion of the solar system and pulsar due to
their orbits around the Galaxy. For the non-globular-cluster pulsars, if their proper motions and
distances are well enough measured, then these effects can be corrected for to give the intrinsic
period derivative \citep[see, e.g.,][]{1991ApJ...366..501D}. For pulsars where the intrinsic period
derivative is given in the literature we have used those values (see Tables~\ref{tab:highvalue} and
\ref{tab:results} for the values and associated references). For further non-globular-cluster
pulsars for which a transverse velocity and distance are given in the ATNF Pulsar Catalog, we
correct the observed period derivative using the method in \citet{1991ApJ...366..501D}. In some
cases the corrections lead to negative period derivative values, indicating that the true values are
actually too small to be confidently constrained. For these cases Table~\ref{tab:results} does not
give a period derivative value or associated spin-down limit.

As was previously done in \citet{2017ApJ...839...12A}, for two globular cluster pulsars,
J1823\textminus3021A and J1824\textminus2452A, we assume that the observed spin-down is not
significantly contaminated by cluster effects following the discussions in
\citet{2011Sci...334.1107F} and \citet{2013ApJ...778..106J}, respectively, so these values are used
without any correction. For the other globular cluster pulsars, we again take the approach of
\citet{2014ApJ...785..119A} and \citet{2017ApJ...839...12A} and create proxy period derivative
values by assuming that the stars have characteristic ages of $10^9$\,yr and braking indices of
$n=5$ (i.e., they are braked purely by gravitational radiation from the $l=m=2$ mode).\footnote{The
braking index $n$ defines the power-law relation between the pulsar's frequency and frequency
derivative via $\dot{f} = -kf^n$, where $k$ is a constant. Purely magnetic dipole braking gives a
value of $n=3$, and purely quadrupole \gw braking gives $n=5$. The characteristic age is defined as
${\tau=(n-1)^{-1}(f/\dot{f})}$.}

\subsubsection{Orientation constraints}\label{sec:orientation}

In \citet{Ng:2004} and \citet{Ng:2008} models are fitted to a selection of X-ray observations of
pulsar wind nebulae, which are used to provide the orientations of the nebulae. In previous \gw
searches \citep{2008ApJ...683L..45A,2010ApJ...713..671A,2014ApJ...785..119A,2017ApJ...839...12A} the
assumption has been made that the orientation of the wind nebula is consistent with the orientation
of its pulsar. In this work we will also follow this assumption and use the fits in \citet{Ng:2008}
as prior constraints on orientation (inclination angle $\iota$ and polarization angle $\psi$) for
PSR J0205+6449, PSR J0534+2200, PSR J0835\textminus4510, PSR J1952+3252, and PSR J2229+6114. This is
discussed in more detail in Appendix~\ref{ap:priors}. We refer to results based on these constraints
as using restricted priors.

Constraints on the position angle, and therefore \gw polarization angle, of pulsars are also
possible through observations of their electromagnetic polarization \citep{2005MNRAS.364.1397J}. None of the pulsars in \citet{2005MNRAS.364.1397J} are in our target
list, but such constraints may be useful in the future. Constraints on the polarization angle alone
are not as useful as those that also provide the inclination of the source (as described above for
the pulsar wind nebula observations), which is directly correlated with the \gw amplitude.
However, there are some pulsars for which double pulses are observed \citep{2008MNRAS.390...87K,
2010MNRAS.402..745K}, suggesting that the rotation axis and magnetic axis are orthogonal, and
therefore implying an inclination angle of $\iota \approx \pm 90^\circ$. In terms of upper limits on
the \gw amplitude, the implication of $\iota \approx 90^\circ$ would generally be to lead to a
larger limit on $h_0$ than for an inclination aligned with the line of sight, due to the relatively
weaker observed strain for a linearly polarized signal compared to a circularly polarized signal of
the same $h_0$. Of the pulsars observed in \citet{2010MNRAS.402..745K}, one (PSR
J1828\textminus1101) is in our search, although we have not used the implied constraints
in this analysis. In the future these constraints will be considered if appropriate.

\section{Results}

For each pulsar the results presented here are from analyses coherently combining the data from both
the LIGO detectors. As described below, we see no strong evidence for a \gw signal from any pulsar,
so we therefore cast our results in terms of upper limits on the \gw amplitude. These limits are
subject to the uncertainties from the detector calibration as described in Section~\ref{sec:gwdata},
as well as statistical uncertainties that are dependent on the particular analysis method used. For
the {\it Bayesian} analysis, statistical uncertainties on the 95\% credible upper limits are on the
order of 1\% \citep[see Figure~12 of][]{2017arXiv170508978P}. For the $5n$-{\it vector} method the
statistical uncertainty on the upper limits is of the order of 1-5\%, depending on the pulsar.

For all pulsars, we present the results of our analyses in terms of several quantities. For the
searches including data at both once and twice the rotation frequency and searching for a signal
from both the $l=2$, $m=1, 2$ modes we present the inferred limits on the $C_{21}$ and $C_{22}$
amplitude parameters given in Equations~(\ref{eq:l2m1}) and (\ref{eq:l2m2}). For the searches
looking only for emission from the $l=m=2$ mode we present limits on the signal's \gw strain $h_0$.
For the {\it Bayesian search} these limits are 95\% credible upper bounds derived from the posterior
probability distributions. For the $5n$-{\it vector} pipeline the upper limits are obtained with a
hybrid frequentist/Bayesian approach, described in Appendix \ref{ap:5vecul}, consisting in
evaluating the posterior probability distribution of the signal amplitude $H_0$, conditioned to the
measured value of a detection statistic, and converting it to a 95\% credible upper limit on $h_0$
or $C_{21}$ (see Section~\ref{sec:methods}, Appendix~\ref{ap:5vecampl} and
\citet{2014ApJ...785..119A} for more details.) Upper limits have been computed assuming both flat
and, when information from electromagnetic observation is available, restricted priors on the
polarization parameters, as detailed in Section~\ref{sec:orientation} and Appendix~\ref{ap:priors}.

For the purely $l=m=2$ mode search, we are able to convert these limits into equivalent limits on
several derived quantities. In cases where we have an estimate for the pulsar distance (see
Section~\ref{sec:pulsars} and Tables~\ref{tab:highvalue} and \ref{tab:results}) $h_0$ can be
converted directly into a limit on the $Q_{22}$ mass quadrupole (see Equation~(\ref{eq:q22})). Under
the assumption of a fiducial principal moment of inertia of $I_{zz}^\text{fid} = 10^{38}\,{\rm
kg}\,{\rm m}^2$ this can also place a limit on the fiducial ellipticity $\varepsilon$. When we also
have a reliable estimate of the intrinsic period derivative, the spin-down limit $h_0^{\rm sd}$ can
be calculated (see Equation~(\ref{eq:h0sd})) and the ratio of the observed limits on $h_0$ to this
value, $h_0^{95\%}/h_0^{\rm sd}$, is shown (the square of this value gives the ratio of the limit on
the \gw luminosity to the spin-down luminosity of the pulsar).

For the {\it Bayesian} method, an odds value giving a ratio of probabilities is also calculated (the
base-10 logarithm of which we denote as $\mathcal{O}$, which is equivalent to
$\log{}_{10}{\mathcal{O}_{\text{S}/\text{I}}}$ from \citealp{2017ApJ...839...12A}), where the
numerator is the probability of the data being consistent with a coherent signal model in both
detectors and the denominator is the probability of an incoherent signal present in both detectors
{\it or} Gaussian noise in one detector and a signal in the other {\it or} Gaussian noise being
present in both detectors (see Appendix A.3 in \citealp{2017ApJ...839...12A} or Section~2.6 of
\citealp{2017arXiv170508978P} for more details). These odds can be used to assess when the coherent
signal model is favored by the data. The values of $\mathcal{O}$ for each pulsar are shown in
Tables~\ref{tab:highvalue} (where it is the value given in the ``Statistic'' column for the {\it
Bayesian} search) and \ref{tab:results}, but in all cases the values are negative, indicating no
pulsars for which the coherent signal model is favored. Also, examination of the posterior
probability distributions for the amplitude parameters shows that none are significantly disjoint
from the probability of the amplitude being zero.

In the $5n$-{\it vector} search the significance of each analysis is expressed through a $p$-value,
which is a measure of how compatible the data are with pure noise. It is obtained by empirically
computing the noise-only distribution of the detection statistic, over an off-source region, and
comparing it to the value of the detection statistic found in the actual analysis. Conventionally, a
threshold of $p < 0.01$ on the $p$-value is used to identify potentially interesting candidates: pulsars
for which the analysis provides a $p$-value smaller than the threshold would deserve a deeper study
\citep[see also][]{2014ApJ...785..119A,2017ApJ...839...12A}. The computed $p$-values are reported in
Table~\ref{tab:highvalue}. For all the analyzed pulsars they are well above $p = 0.01$,
suggesting that the data are fully compatible with noise.

For the $\mathcal{F}$-/$\mathcal{G}$-{\it statistic} method false-alarm probabilities of obtaining
the observed statistic values are calculated. They are derived assuming that for the
$\mathcal{F}$-statistic the $2\mathcal{F}$ value has a $\chi^2$ distribution with 4
degrees of freedom \citep{1998PhRvD..58f3001J} and for the $\mathcal{G}$-statistic the
$2\mathcal{G}$ value has a $\chi^2$ distribution with 2 degrees of freedom \citep{Jaranowski:2010}.
The false-alarm probabilities reported in Table~\ref{tab:highvalue} are all close to unity and show
no strong indication that the statistics deviate from their expected distributions.

\begin{figure*}
\epsscale{1.15}
\plotone{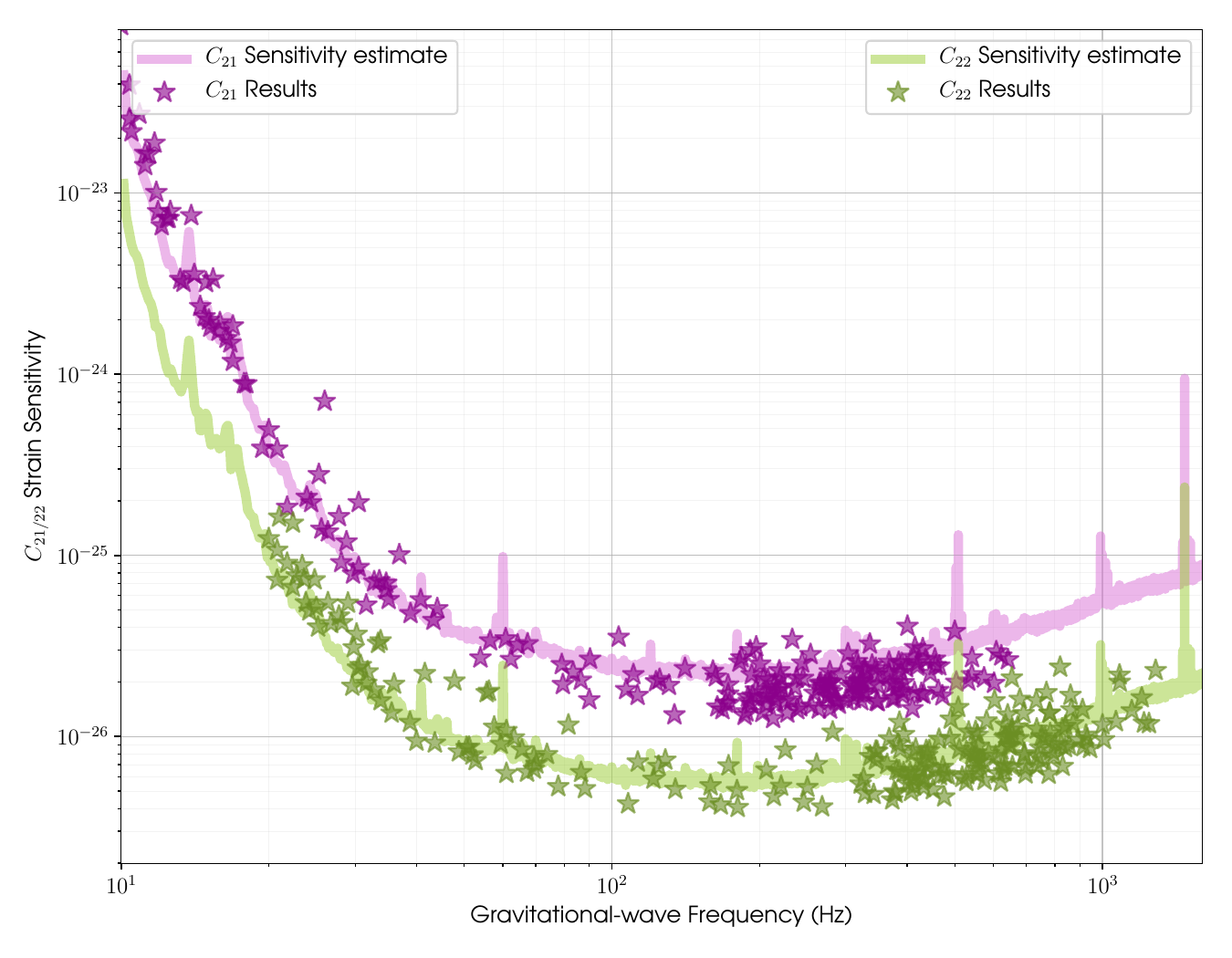}
\caption{Upper limits on $C_{21}$ and $C_{22}$ for \NPULSARS pulsars. The stars show the observed 95\% credible upper limits on observed amplitudes for each pulsar. The solid lines show an estimate of the expected sensitivity of the searches.\label{fig:resultsC}}
\end{figure*}

\begin{figure*}
\epsscale{1.15}
\plotone{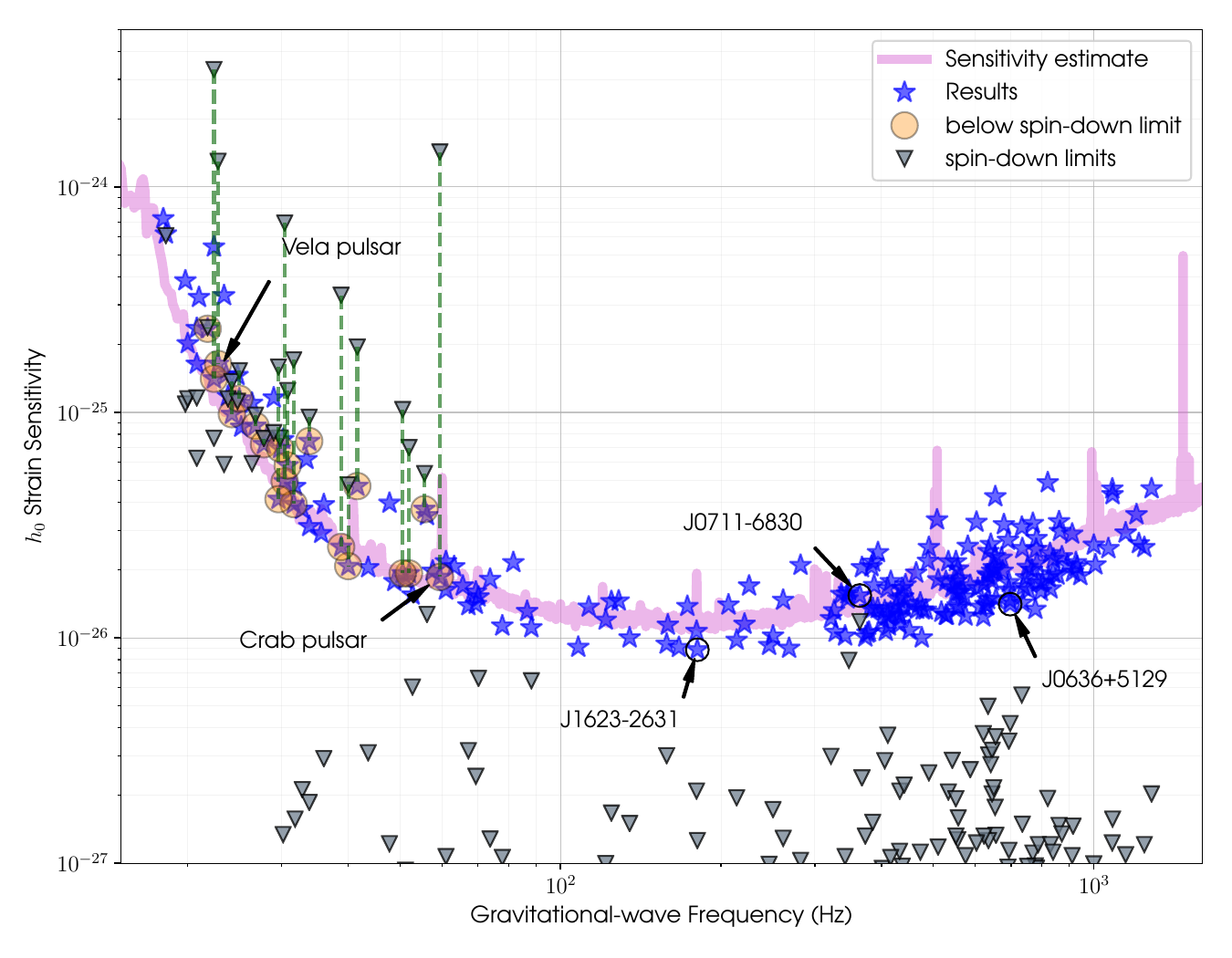}
\caption{Upper limits on $h_0$ for \NPULSARS pulsars. The stars show the observed 95\% credible upper limits on observed amplitude for each pulsar. The solid line shows an estimate of the expected sensitivity of the search. Triangles show the limits on \gw amplitude derived from each pulsar's observed spin-down.\label{fig:resultsh0}}
\end{figure*}

The results for the \NHIGHVALUE high-value targets are shown in Table~\ref{tab:highvalue} and
the results for all the other pulsars are shown in Table~\ref{tab:results}. The 95\% credible upper
limits on $C_{21}$ and $C_{22}$ for all \NPULSARS pulsars from the {\it Bayesian} analysis are shown
as a function of the \gw emission frequency in Figure~\ref{fig:resultsC}. Also shown are estimates
of the expected sensitivity of the search given representative noise amplitude spectral densities
from the O1 and O2 observing runs (see Appendix~\ref{ap:sensitivity} for descriptions of how these
were produced). The 95\% credible upper limits on $h_0$ for all \NPULSARS pulsars from the search
purely for emission from the $l=m=2$ mode are shown in Figure~\ref{fig:resultsh0}.
Figure~\ref{fig:resultsh0} also shows spin-down limits on the emission as dark triangles, and in the
cases where our observed upper limits are below these the result is highlighted with a circular
marker and is linked to its associated spin-down limit with a vertical line.

\begin{figure}
\epsscale{1.15}
\plotone{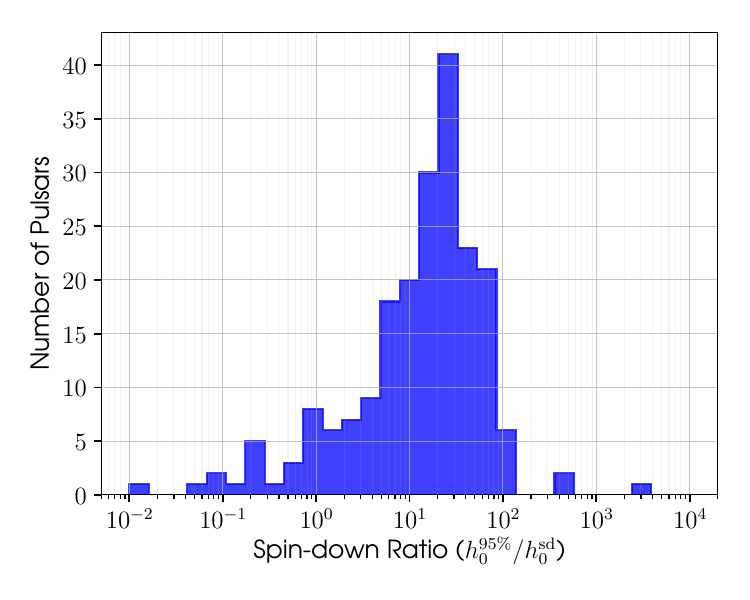}
\caption{Histogram of ratios of upper limits on $h_0$ compared to the spin-down limit.\label{fig:sdratio}}
\end{figure}

Figure~\ref{fig:sdratio} shows a histogram of the spin-down ratio $h_0^{95\%}/h_0^{\text{sd}}$ from
the {\it Bayesian} analysis for the $l=m=2$ mode search, for pulsars where it was possible to
calculate a spin-down limit. This shows \NBELOWSPINDOWN pulsars for which $h_0^{95\%} <
h_0^{\text{sd}}$, and \NSPINDOWNONETOTEN for which the results are between 1 and 10 times greater
than $h_0^{\text{sd}}$. If we just look at \acp{MSP}, then \NSPINDOWNBELOWTENMSP are within a factor
of 10 of the spin-down limit.\footnote{Based on our sample of pulsars with rotation frequencies
greater than 10\,Hz, there is a clear distinction between the \ac{MSP} and young (or normal)
population based on a cut in $\dot{P}$ of $10^{-17}$\,s\,s$^{-1}$, i.e., we assume that any pulsar
with a $\dot{P}$ smaller than this is an \ac{MSP}.} The spin-down limits and the $Q_{22}$ and
$\varepsilon$ values assume a particular distance, intrinsic period derivative, and fiducial moment
of inertia of $10^{38}\,\text{kg}\,\text{m}^2$, but there can be considerable uncertainties on these
values. For example, distances calculated using the Galactic electron density model of
\citet{2017ApJ...835...29Y} have a $1\sigma$ relative error of $\sim 40\%$, with some parts of the
sky having several 100\% relative errors. The true moment of inertia depends on the pulsar's mass
and equation of state and could be within a range of roughly
$(1\text{--}3)\!\times\!10^{38}\,\text{kg}\,\text{m}^2$ (see, e.g., Figures~4 and 7 of
\citealt{2008ApJ...685..390W} and Figures~6 and 7 of \citealt{2013A&A...552A..59B}). We do not
incorporate these uncertainties into the results we present here, but they should be kept in mind
when interpreting the limits.\footnote{From Equations~(\ref{eq:ell}), (\ref{eq:q22}), and
(\ref{eq:h0sd}) it can be seen that fractional uncertainties on distance will scale directly into
the uncertainties on $\varepsilon$, $Q_{22}$ and $h_0^{\text{sd}}$. Increasing the value of
$I_{zz}^{\text{fid}}$ will proportionally decrease the inferred $\varepsilon$ value and increase
the inferred spin-down limit by a factor given by the square root of the fractional increase
compared to the canonical moment of inertia.} In the case of pulsar distances the references
provided in Tables~\ref{tab:highvalue} and \ref{tab:results} should be consulted to provide an
estimate of the associated uncertainty. These uncertainties dominate the few percent uncertainties
arising from the calibration of the \gw detectors described in Section~\ref{sec:gwdata}.

\begin{figure*}
\epsscale{1.15}
\plotone{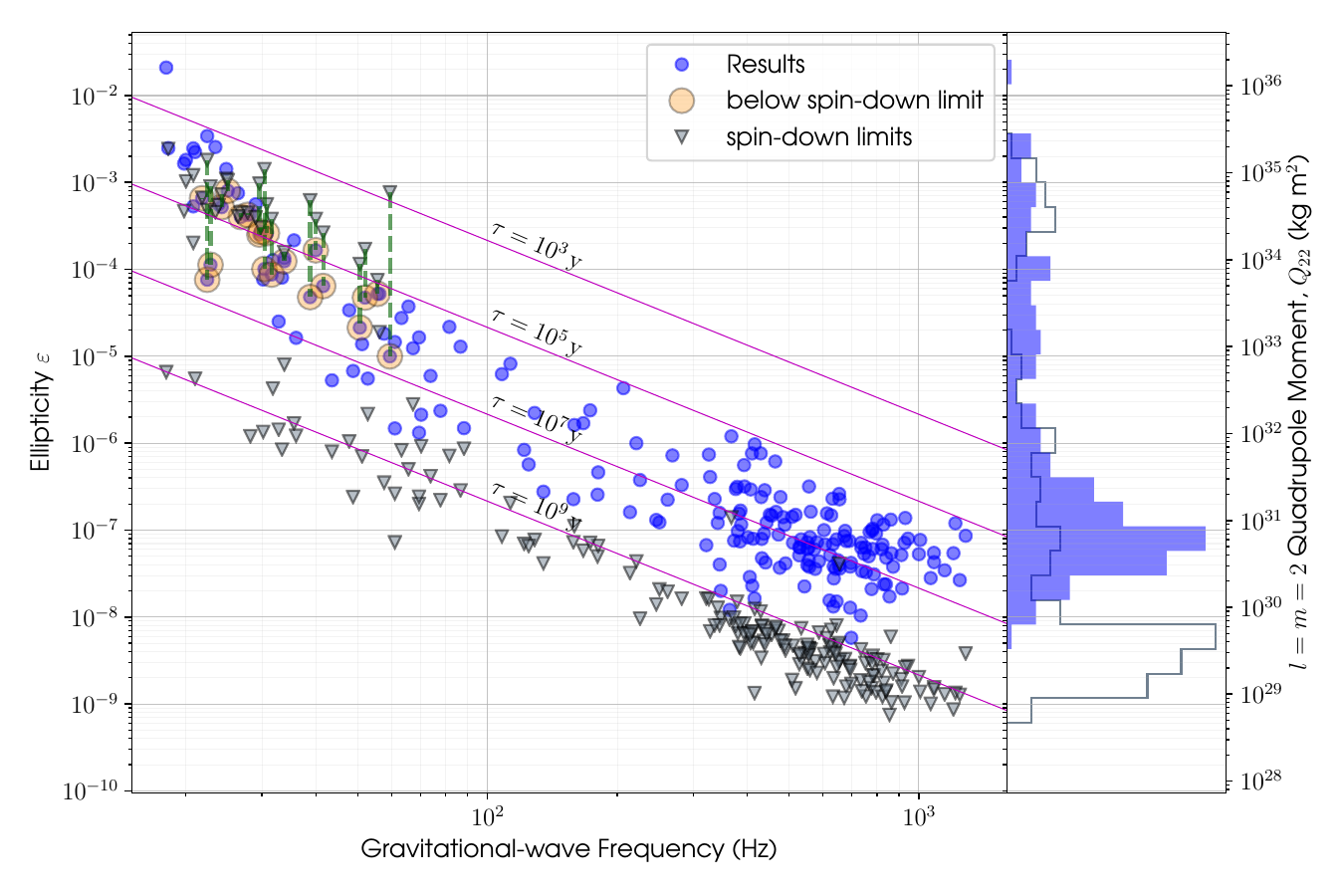}
\caption{Upper limits on mass quadrupole $Q_{22}$ and fiducial ellipticity $\varepsilon$ for
\NPULSARS pulsars. The filled circles show the limits as derived from the observed upper limits on
the \gw amplitude $h_0$ assuming the canonical moment of inertia and distances given in
Tables~\ref{tab:highvalue} and \ref{tab:results}. Triangles show the limits based derived from each
pulsar's observed spin-down. The diagonal lines show contours of equal characteristic age $\tau$
assuming that braking is entirely through \gw emission. The distributions of these limits are also
show in histogram form to the right of the figure, with the filled and unfilled histograms showing
our observed limits and the spin-down limits, respectively.\label{fig:resultsell}}
\end{figure*}

The $h_0^{95\%}$ results from the {\it Bayesian} analysis, recast as limits on $Q_{22}$ and the
fiducial ellipticity and assuming the distances given in Tables~\ref{tab:highvalue} and
\ref{tab:results}, are shown in Figure~\ref{fig:resultsell}. The much lower limits on $\varepsilon$
inferred for the \acp{MSP} easily follow from the frequency scaling seen in
Equation~(\ref{eq:fidell}).

\subsection{Results highlights}

For decades, two of the most intriguing targets in searches for gravitational waves from pulsars
have been the Crab and Vela pulsars (J0534+2200 and J0835\textminus4510, respectively), due to their
large spin-down luminosities. For these two pulsars, assuming emission from the $l=m=2$ mode
and with the phase precisely locked to the observed rotational phase, the limits observed
using the initial LIGO and Virgo detectors in \citet{2008ApJ...683L..45A} and
\citet{2011ApJ...737...93A}, respectively, were lower than the equivalent spin-down limits.
Using data from the O1 run, the observed limits were also below the spin-down limit for
these two pulsars in searches where the strict phase locking of the observed rotational phase and
\gw phase was relaxed \citep{2017PhRvD..96l2006A}.\footnote{In the similar narrowband
searches for the Crab pulsar in \citet{2008ApJ...683L..45A} and \citet{2015PhRvD..91b2004A} the
limits were also below the spin-down limit, under the assumption that the orientation was
restricted to that derived from the pulsar wind nebula (see Section~\ref{sec:orientation}).}

For the Crab pulsar, this analysis finds an observed 95\% limit of $h_0^{95\%} = \CRABHZEROBAYES$
for the {\it Bayesian} analysis (with consistent values of \CRABHZEROFSTAT and \CRABHZEROFIVENVEC
for the $\mathcal{F}$-{\it statistic} and 5$n$-{\it vector} analyses, respectively). This is
\CRABSDBAYES times the spin-down ratio, or, equivalently, it means that less than \CRABPOWERBAYES of
the available spin-down luminosity is emitted via gravitational waves (see
Equation~(\ref{eq:luminosity})). These limits are also well below less naive spin-down limits that
can be calculated by taking into account the power radiated electromagnetically or through particle
acceleration \citep{1969ApJ...157.1395O,2000A&A...354..163P}. As shown in Table~\ref{tab:highvalue},
slightly tighter constraints are possible if one assumes that the orientation of the pulsar matches
that derived from the observed orientation of its pulsar wind nebula (see
Section~\ref{sec:orientation}). The above $h_0$ upper limit corresponds to limits on $Q_{22}$ of
\CRABQTTBAYES and an equivalent fiducial ellipticity of \CRABELLIPTICITYBAYES. This mass quadrupole
is almost in the range of maximum allowable quadrupoles for standard neutron star equations of state
(see discussion in Section~\ref{sec:strength} and \citealp{2013PhRvD..88d4004J}).

Similarly, for the Vela pulsar, this analysis finds an observed 95\% limit of $h_0^{95\%} =
\VELAHZEROBAYES$ for the {\it Bayesian} analysis (with broadly consistent values of \VELAHZEROFSTAT
and \VELAHZEROFIVENVEC for the $\mathcal{F}$-{\it statistic} and 5$n$-{\it vector} analyses,
respectively). This is \VELASDBAYES times the spin-down ratio, or, equivalently, means that less
than \VELAPOWERBAYES of the available spin-down luminosity is emitted via gravitational waves. The
above $h_0$ upper limit corresponds to limits on $Q_{22}$ of \VELAQTTBAYES and an equivalent
fiducial ellipticity of \VELAELLIPTICITYBAYES.

Of all the pulsars in the analysis, the one with the smallest upper limit on $h_0$ is
PSR~\SMALLESTHZEROPSR (with a rotational frequency of \SMALLESTHZEROFREQ and distance of
\SMALLESTHZERODIST), with $h_0^{95\%} = \SMALLESTHZERO$. The pulsar with the smallest limit on the
$Q_{22}$ mass quadrupole is PSR~\SMALLESTQTTPSR (with a rotational frequency of \SMALLESTQTTFREQ and
distance of \SMALLESTQTTDIST), with $Q_{22}^{95\%}$ of \SMALLESTQTT, and an equivalent fiducial
ellipticity limit of \SMALLESTELLIPTICITY. These limits are only a factor of \SMALLESTQTTRATIO above
the pulsar's spin-down limit. Of the \acp{MSP} in our search (which, as above, we take as any pulsar
with $\dot{P} < 10^{-17}$\,s\,s$^{-1}$), the one for which our limit is closest to the spin-down
limit is \SMALLESTMSPSDRATPSR (with a rotational frequency of \SMALLESTMSPSDRATFREQ and a distance
of \SMALLESTMSPSDRATDIST). It is within a factor of \SMALLESTMSPSDRAT of the spin-down limit, with
an observed upper limit of $h_0^{95\%} = \SMALLESTMSPSDRATHZERO$ and derived limits on $Q_{22}$ and
ellipticity of \SMALLESTMSPSDRATQTT and \SMALLESTMSPSDRATELLIPTICITY, respectively.\footnote{It is
interesting to note that in \citet{2017ApJ...839...12A} PSR~J0437\textminus4715 was the \ac{MSP}
with an observed upper limit closest to its spin-down limit, being only a factor of 1.4 above that
value, while \SMALLESTMSPSDRATPSR had a limit that was a factor of $\sim 20$ above its spin-down
limit. For J0437\textminus4715, despite now having an improved upper limit on the \gw amplitude, the
correction of the observed period derivative to the intrinsic period derivative has lowered the
spin-down limit by roughly a factor of two. For \SMALLESTMSPSDRATPSR the distance estimated using
the YMW16 Galactic electron density model \citep{2017ApJ...835...29Y} is about a factor of 9 closer
than that estimated with the previously used NE2001 model \citep{2002astro.ph..7156C}.} The upper
bound on possible neutron star moments of inertia is roughly $\scinum{3}{38}$\,kg\,m$^2$, for which
the fiducial spin-down limit could be increased by a factor of $\sqrt{3} \approx 1.7$, which would
be greater than our upper limit.

Similarly to \citet{2017ApJ...839...12A}, our most stringent limits on ellipticity for \acp{MSP}
still imply limits on the internal toroidal magnetic field strength of $\lesssim 10^{9}$\,T (or
$10^{13}$\,G) \citep[applying Equation (2.4) of][and assuming a superconducting
core]{2002PhRvD..66h4025C}. The method in \citet{2012MNRAS.421..760M} could also be applied to
these results to constrain the ratio of the poloidal magnetic field energy to the total field
energy.

For the searches that include the $l=2$, $m=1$ mode, the smallest upper limit on the $C_{21}$
amplitude is for PSR~\SMALLESTCTOPSR (with a rotational frequency of \SMALLESTCTOFREQ), at
$C_{21}^{95\%} = \SMALLESTCTO$. As $C_{21}$ and $C_{22}$ are not very strongly correlated, the upper
limits on $C_{22}$ are generally consistent with $C_{22}^{95\%} \approx h_0^{95\%}/2$.

\clearpage
\startlongtable
\movetabledown=2.4cm
\begin{longrotatetable}
\begin{deluxetable}{lrllllllllllll}
\tablecaption{Limits on Gravitational-wave Amplitude, and Other Derived Quantities, for 34 High-value Pulsars from the Three Analysis Methods.\label{tab:highvalue}}
\tabletypesize{\footnotesize}
\tablehead{
\colhead{Pulsar Name} & 
\colhead{$f_{\text{rot}}$} & 
\colhead{$\dot{P}_{\text{rot}}$} & 
\colhead{Distance} & 
\colhead{$h_0^{\text{sd}}$} & 
\colhead{Analysis} & 
\colhead{$C_{21}^{95\%}$} & 
\colhead{$C_{22}^{95\%}$} & 
\colhead{$h_0^{95\%}$} & 
\colhead{$Q_{22}^{95\%}$} & 
\colhead{$\varepsilon^{95\%}$} & 
\colhead{$h_0^{95\%}/h_0^{\text{sd}}$} & 
\colhead{Statistic\tablenotemark{a}} & 
\colhead{Statistic\tablenotemark{b}}\\ 
\colhead{(J2000)} & 
\colhead{(Hz)} & 
\colhead{(s\,s$^{-1})$} & 
\colhead{(kpc)} & 
\colhead{~} & 
\colhead{Method} & 
\colhead{~} & 
\colhead{~} & 
\colhead{~} & 
\colhead{(kg\,m$^2$)} & 
\colhead{~} & 
\colhead{~} & 
\colhead{$^{l=2, m=1,2}$} & 
\colhead{$^{l=2, m=2}$}}
\startdata
\multirow{3}{*}{J0030+0451} & \multirow{3}{*}{205.5} & \multirow{3}{*}{\scinum{1.1}{-20} (g)} & \multirow{3}{*}{0.33 (a)} & \multirow{3}{*}{\scinum{3.7}{-27}} & Bayesian & \scinum{1.7}{-26} & \scinum{5.9}{-27} & \scinum{1.3}{-26} & \scinum{1.8}{30} & \scinum{2.3}{-8} & 3.4 & -3.8 & -2.1 \\ 
 &  &  &  &  & $\mathcal{F}$-statistic & \nodata & \nodata & \nodata & \nodata & \nodata & \nodata & \nodata & \nodata \\ 
 &  &  &  &  & 5$n$-vector & \scinum{1.3}{-26} & \nodata & \scinum{1.7}{-26} & \scinum{2.3}{30} & \scinum{3.0}{-8} & 4.5 & 0.72 & 0.61 \\ 
\hline 
\multirow{3}{*}{J0117+5914\tablenotemark{c}} & \multirow{3}{*}{9.9} & \multirow{3}{*}{\scinum{5.9}{-15}} & \multirow{3}{*}{1.77 (b)} & \multirow{3}{*}{\scinum{1.1}{-25}} & Bayesian & \nodata & \nodata & \scinum{3.8}{-25} & \scinum{1.3}{35} & \scinum{1.7}{-3} & 3.5 & -2.4 & -1.9 \\ 
 &  &  &  &  & $\mathcal{F}$-statistic & \nodata & \nodata & \nodata & \nodata & \nodata & \nodata & \nodata & \nodata \\ 
 &  &  &  &  & 5$n$-vector & \nodata & \nodata & \scinum{2.6}{-25} & \scinum{8.6}{34} & \scinum{1.1}{-3} & 2.4 & \nodata & 0.31 \\ 
\hline 
\multirow{3}{*}{J0205+6449\tablenotemark{c}} & \multirow{3}{*}{15.2} & \multirow{3}{*}{\scinum{1.9}{-13}} & \multirow{3}{*}{2.00 (c)} & \multirow{3}{*}{\scinum{6.9}{-25}} & Bayesian & \scinum{1.8(1.5)}{-24} & \scinum{2.4(3.6)}{-26} & \scinum{4.9(7.1)}{-26} & \scinum{0.8(1.1)}{33} & \scinum{1.0(1.5)}{-4} & 0.071(0.1) & -4.8(-4.6) & -2.7(-2.4) \\ 
 &  &  &  &  & $\mathcal{F}$-statistic & \scinum{2.2}{-24} & \scinum{4.5}{-26} & \scinum{8.8}{-26} & \scinum{1.4}{34} & \scinum{1.8}{-4} & 0.13 & 0.71 & 0.26 \\
\enddata
\tablecomments{For references and other notes see Table~\ref{tab:results}. Values in parentheses are those produced using the restricted orientation priors described in Section~\ref{sec:orientation}.}
\tablenotetext{a}{For the {\it Bayesian} method this column shows the base-10 logarithm of the Bayesian odds, $\mathcal{O}$, comparing a coherent signal model at both the $l=2$, $m=1,2$ modes to incoherent signal models. For the $\mathcal{F}$-/$\mathcal{G}$-{\it statistic} method this column shows the false-alarm probability for a signal just at the $l=2$, $m=1$ mode, assuming that the $2\mathcal{F}$ value has a $\chi^2$ distribution with 4 degrees of freedom and the $2\mathcal{G}$ value has a $\chi^2$ distribution with 2 degrees-of-freedom. For the 5$n$-{\it vector} method this column shows the $p$-value for a search for a signal at just the $l=2$, $m=1$ mode, where the null hypothesis being tested is that the data are consistent with pure Gaussian noise.}
\tablenotetext{b}{This is the same as in footnote \tablenotemark{a}, but for all the methods the assumed signal model is from the $l=m=2$ mode.}
\tablenotetext{c}{The observed $\dot{P}$ has been corrected to account for the relative motion between the pulsar and observer.}
\end{deluxetable}
\end{longrotatetable}

\clearpage
\startlongtable
\begin{longrotatetable}
\begin{deluxetable}{lrlllhllllllll}
\tablecaption{Limits on Gravitational-wave Amplitude and Other Derived Quantities for 187  Pulsars from the {\it Bayesian} Analysis.\label{tab:results}}
\tabletypesize{\footnotesize}
\tablehead{
\colhead{Pulsar Name} & 
\colhead{$f_{\text{rot}}$} & 
\colhead{$\dot{P}_{\text{rot}}$} & 
\colhead{Distance} & 
\colhead{$h_0^{\text{sd}}$} & 
\nocolhead{Analysis} & 
\colhead{$C_{21}^{95\%}$} & 
\colhead{$C_{22}^{95\%}$} & 
\colhead{$h_0^{95\%}$} & 
\colhead{$Q_{22}^{95\%}$} & 
\colhead{$\varepsilon^{95\%}$} & 
\colhead{$h_0^{95\%}/h_0^{\text{sd}}$} & 
\colhead{$\mathcal{O}^{l=2}_{m=1,2}$} & 
\colhead{$\mathcal{O}^{l=2}_{m=2}$}\\ 
\colhead{(J2000)} & 
\colhead{(Hz)} & 
\colhead{(s\,s$^{-1})$} & 
\colhead{(kpc)} & 
\colhead{~} & 
\nocolhead{Method} & 
\colhead{~} & 
\colhead{~} & 
\colhead{~} & 
\colhead{(kg\,m$^2$)} & 
\colhead{~} & 
\colhead{~} & 
\colhead{~} & 
\colhead{~}}
\startdata
J0023+0923\tablenotemark{a} & 327.8 & \scinum{1.0}{-20} & 1.10 (a) & \scinum{1.3}{-27} & Bayesian & \scinum{2.4}{-26} & \scinum{6.8}{-27} & \scinum{1.5}{-26} & \scinum{2.8}{30} & \scinum{3.6}{-8} & 11 & -3.9 & -2.2 \\ 
J0034\textminus0534\tablenotemark{a} & 532.7 & \scinum{4.2}{-21} & 1.35 (b) & \scinum{8.9}{-28} & Bayesian & \scinum{2.0}{-26} & \scinum{1.2}{-26} & \scinum{2.5}{-26} & \scinum{2.2}{30} & \scinum{2.8}{-8} & 28 & -4.1 & -2.1 \\ 
J0101\textminus6422\tablenotemark{a} & 388.6 & \scinum{3.8}{-21} & 1.00 (b) & \scinum{9.7}{-28} & Bayesian & \scinum{2.3}{-26} & \scinum{6.2}{-27} & \scinum{1.3}{-26} & \scinum{1.6}{30} & \scinum{2.1}{-8} & 14 & -4.1 & -2.3 \\ 
J0102+4839 & 337.4 & \scinum{1.1}{-20} & 2.38 (b) & \scinum{6.6}{-28} & Bayesian & \scinum{1.9}{-26} & \scinum{9.8}{-27} & \scinum{2.0}{-26} & \scinum{7.6}{30} & \scinum{9.8}{-8} & 30 & -4.0 & -1.9 \\ 
J0218+4232\tablenotemark{a} & 430.5 & \scinum{7.7}{-20} & 3.15 (d) & \scinum{1.5}{-27} & Bayesian & \scinum{3.1}{-26} & \scinum{1.7}{-26} & \scinum{3.3}{-26} & \scinum{1.0}{31} & \scinum{1.3}{-7} & 22 & -3.0 & -1.7 \\ 
J0248+4230 & 384.5 & \scinum{1.7}{-20} & 1.85 (b) & \scinum{1.1}{-27} & Bayesian & \scinum{2.6}{-26} & \scinum{1.8}{-26} & \scinum{3.2}{-26} & \scinum{7.4}{30} & \scinum{9.5}{-8} & 29 & -3.4 & -1.8 \\ 
J0251+26 & 393.5 & \scinum{7.6}{-21} & 1.15 (b) & \scinum{1.2}{-27} & Bayesian & \scinum{2.0}{-26} & \scinum{8.4}{-27} & \scinum{1.8}{-26} & \scinum{2.4}{30} & \scinum{3.1}{-8} & 15 & -4.0 & -2.1 \\ 
J0308+74 & 316.8 & \scinum{1.7}{-20} & 0.38 (b) & \scinum{5.0}{-27} & Bayesian & \scinum{1.7}{-26} & \scinum{6.9}{-27} & \scinum{1.5}{-26} & \scinum{1.0}{30} & \scinum{1.3}{-8} & 3 & -3.9 & -2.2 \\
\enddata
\tablenotetext{a}{The observed $\dot{P}$ has been corrected to account for the relative motion between the pulsar and observer.}
\tablenotetext{b}{The corrected pulsar $\dot{P}$ value is negative, so no value is given and no spin-down limit has been calculated.}
\tablenotetext{c}{This is a globular cluster pulsar for which a proxy period derivative has been derived assuming a characteristic age of $10^9$\,years and a braking index of $n=5$.}
\tablerefs{The following is a list of references for pulsar distances and intrinsic period derivatives, and they should be consulted for information on the associated uncertainties on these quantities: (a) \citet{2018ApJS..235...37A}, (b) \citet{2017ApJ...835...29Y}, (c) \citet{2013AA...560A..18K}, (d) \citet{2014MNRAS.444.1859V}, (e) \citet{2013Sci...340..448A}, (f) \citet{2016MNRAS.455.1751R}, (g) \citet{2016MNRAS.458.3341D}, (h) \citet{2016MNRAS.455.3806B}, (i) \citet{2009Sci...323.1327D}, (j) \citet{2003ApJ...596.1137D}, (k) Mingarelli, private communication, (l) \citet{2017ApJ...839...12A}, (m) \citet{2012ApJ...755...39V}, (n) \citet{2013ApJ...763...80B}, (o) \citet{2013ApJ...778..120H}, (p) \citet{2014ApJ...787...82F}, (q) \citet{2015ApJ...799..165B}, (r) \citet{2018ApJ...855..122V}, (s) \citet{2018arXiv181206262M}, (t) \citet{2012MNRAS.423.3328F}, (u) \citet{2013MNRAS.430..571E}, (v) \citet{2007AA...470.1043O}, (w) \citet{2014MNRAS.443.2183F}, (x) \citet{1996yCat.7195....0H}, (y) \citet{2010MNRAS.402.1729V}, (z) \citet{2014ApJ...795..168M}, (aa) \citet{2007AJ....133.1287V}, (bb) \citet{1991AJ....102..152R}, (cc) \citet{2011RAA....11..824W}, (dd) \citet{2011ApJ...729L..16G}, (ee) \citet{2003AA...408..529G}, (ff) \citet{2004ApJ...602..264M}, (gg) \citet{2013ApJ...770..145D}, (hh) \citet{2001ApJ...547..323H}, (ii) \citet{2018MNRAS.475..469S}, (jj) \citet{2011ApJ...743L..26R}, (kk) \citet{2014MNRAS.439.1865N}.}
\end{deluxetable}
\end{longrotatetable}

\section{Discussion}

In this paper we have used data from the first two observation runs of Advanced LIGO (O1 and O2) to
update the upper limits on the \gw amplitude $h_0$ for emission from the $l=m=2$ mass quadrupole for
\PULSAROVERLAP pulsars. This compares to \NPREVPULSARS results presented previously in
\citet{2014ApJ...785..119A} (using data from the initial runs of the LIGO
\citep{2009RPPh...72g6901A} and Virgo \citep{2012JInst...7.3012A} detectors, S1--6 and VSR1--4) and
\citet{2017ApJ...839...12A} (using data from the first observing run, O1, of the advanced LIGO
detectors; \citealp{2015CQGra..32g4001L, 2016PhRvL.116m1103A}). New upper limits on $h_0$ have been
set for a further \PULSARNEW pulsars. Other than the results in \citet{2015MNRAS.453.4399P}, we have
also presented the first comprehensive set of results for searches that also include the possibility
of emission from the $l=2$, $m=1$ mode at the pulsar's rotation frequency. These are expressed as
upper limits on two amplitude parameters $C_{21}$ and $C_{22}$ defined in
\citet{2015MNRAS.453...53J}. We find no strong evidence for \gw emission from any pulsar in the
searches purely for the $l=m=2$ mode, or both the $l=2$, $m=1,\,2$ modes.

Further analyses of this dataset are possible. For example, we have not presented any updated
results regarding potential emission from nontensorial polarization modes as performed in
\citet{2018PhRvL.120c1104A}. In addition to this, the results from all pulsars could be combined in
a way, such as that described in \citet{2018PhRvD..98f3001P}, to constrain the underlying pulsar
ellipticity distribution and determine if the ensemble of all pulsar provides evidence for any \gw
signal.

With the \acp{MSP} PSR~\SMALLESTQTTPSR and PSR \SMALLESTMSPSDRATPSR within a factor of $\sim 3$ of their
respective spin-down limits, the imminent third observing run of the advanced LIGO and Virgo
detectors (O3) could allow us to obtain limits below the spin-down limit for an \ac{MSP} for the first time.
This offers the intriguing possibility for signal detection from these extremely smooth objects,
with spin-down-derived ellipticities of a few $10^{-9}$. The O3 sensitivity could also bring the
limits for the Crab pulsar into the range of mass quadrupoles allowed by reasonably standard neutron
star equations of state.

\acknowledgments


The authors gratefully acknowledge the support of the United States National Science Foundation
(NSF) for the construction and operation of the LIGO Laboratory and Advanced LIGO as well as the
Science and Technology Facilities Council (STFC) of the United Kingdom, the Max-Planck-Society
(MPS), and the State of Niedersachsen/Germany for support of the construction of Advanced LIGO  and
construction and operation of the GEO600 detector.  Additional support for Advanced LIGO was
provided by the Australian Research Council. The authors gratefully acknowledge the Italian Istituto
Nazionale di Fisica Nucleare (INFN), the French Centre National de la Recherche Scientifique (CNRS)
and the Foundation for Fundamental Research on Matter supported by the Netherlands Organisation for
Scientific Research, for the construction and operation of the Virgo detector and the creation and
support  of the EGO consortium. The authors also gratefully acknowledge research support from these
agencies as well as by the Council of Scientific and Industrial Research of India, the Department of
Science and Technology, India, the Science \& Engineering Research Board (SERB), India, the Ministry
of Human Resource Development, India, the Spanish Agencia Estatal de Investigaci\'on, the
Vicepresid\`encia i Conselleria d'Innovaci\'o, Recerca i Turisme and the Conselleria d'Educaci\'o i
Universitat del Govern de les Illes Balears, the Conselleria d'Educaci\'o, Investigaci\'o, Cultura i
Esport de la Generalitat Valenciana, the National Science Centre of Poland, the Swiss National
Science Foundation (SNSF), the Russian Foundation for Basic Research, the Russian Science
Foundation, the European Commission, the European Regional Development Funds (ERDF), the Royal
Society, the Scottish Funding Council, the Scottish Universities Physics Alliance, the Hungarian
Scientific Research Fund (OTKA), the Lyon Institute of Origins (LIO), the National Research,
Development and Innovation Office Hungary (NKFI),  the National Research Foundation of Korea,
Industry Canada and the Province of Ontario through the Ministry of Economic Development and
Innovation, the Natural Science and Engineering Research Council Canada, the Canadian Institute for
Advanced Research, the Brazilian Ministry of Science, Technology, Innovations, and Communications,
the International Center for Theoretical Physics South American Institute for Fundamental Research
(ICTP-SAIFR),  the Research Grants Council of Hong Kong, the National Natural Science Foundation of
China (NSFC), the Leverhulme Trust, the Research Corporation, the Ministry of Science and Technology
(MOST), Taiwan and the Kavli Foundation. The authors gratefully acknowledge the support of the NSF,
STFC, MPS, INFN, CNRS and the State of Niedersachsen/Germany for provision of computational
resources.

The Nan\c{c}ay Radio Observatory is operated by the Paris Observatory, associated with the French
CNRS. We acknowledge financial support from the ``Programme National Gravitation, R\'ef\'erences,
Astronomie, M\'etrologie’’ (PNGRAM) and ``Programme National Hautes \'Energies’’ (PNHE) of
CNRS/INSU, France. Work at the Naval Research Laboratory is supported by NASA. We gratefully
acknowledge the continuing contributions of the NICER science team in providing up-to-date spin
ephemerides for X-ray-bright pulsars of interest to the LVC. NICER is a 0.2--12 keV X-ray telescope
operating on the International Space Station. The NICER mission and portions of the NICER science
team activities are funded by NASA.

This work has been assigned LIGO document number LIGO-P1800344.

\software{Much of the analysis described in the paper was performed using the publicly available {\sc LALSuite} library \citep{LALSuite}. Production of many of the pulsar timing ephemerides used in this analysis was performed with {\sc Tempo}\footnote{\url{http://tempo.sourceforge.net/}} and {\sc Tempo2} \citep{2006MNRAS.369..655H}. Figures in this publication have be produced using Matplotlib \citep{2007CSE.....9...90H}.}

\facilities{Arecibo, Fermi, LIGO, Lovell, Molonglo Observatory, MtPO:26m, NICER, NRT, Parkes}

\appendix

\section{Definitions}\label{ap:definitions}

Here we will define some of the standard useful quantities reported and used in our results
\citep[many of these are defined in][]{2014ApJ...785..119A}. The standard definition for the \gw
amplitude from the $l=m=2$ mass quadrupole for a nonprecessing triaxial star rotating about a
principal axis is
\begin{equation}\label{eq:h0}
h_0 = \frac{16\pi^2 G}{c^4}\frac{I_{zz}^\text{fid}\varepsilon f_{\text{rot}}^2}{d}
\approx \scinum{4.23}{-26}\left(\frac{1\,\text{kpc}}{d}\right)  \left(\frac{I_{zz}^\text{fid}}{10^{38}\,\text{kg}\,\text{m}^2}\right) \left( \frac{\varepsilon}{10^{-6}} \right)\left(\frac{f_{\text{rot}}}{100\,\text{Hz}}\right)^2,
\end{equation}
where $d$ is the pulsar distance, $I_{zz}^\text{fid}$ is the fiducial component of the
moment-of-inertia tensor ellipsoid about the rotation axis, $f_\text{rot}$ is the pulsar's rotation
frequency, and $\varepsilon$ is the star's fiducial ellipticity \citep[see,
e.g.,][]{2013PhRvD..88d4016J} defined as
\begin{equation}\label{eq:ell}
\varepsilon = \frac{|I_{xx}-I_{yy}|}{I_{zz}^\text{fid}},
\end{equation}
where $I_{xx}$ and $I_{yy}$ are the true moments of inertia about the principal axes other than the
rotation axis.

The \gw amplitude is related to the $l=m=2$ mass quadrupole $Q_{22}$ via
\begin{equation}\label{eq:q22}
Q_{22} \equiv I_{zz}^\text{fid}\varepsilon \sqrt{\frac{15}{8\pi}} = h_0 \left(\frac{c^4 d}{16\pi^2G f_\text{rot}^2}\right)  \sqrt{\frac{15}{8\pi}} 
\approx \scinum{1.83}{32}\left(\frac{h_0}{10^{-25}}\right) \left(\frac{d}{1\,\text{kpc}}\right)\left( \frac{100\,\text{Hz}}{f_\text{rot}} \right)^2\,\text{kg}\,\text{m}^2,
\end{equation}
where we use the definition of the mass quadrupole used in \citealp{2005PhRvL..95u1101O} and
defined in \citet{2000MNRAS.319..902U}. Alternatively, we can use $h_0$ to calculate the fiducial
ellipticity, defined as
\begin{equation}\label{eq:fidell}
\varepsilon = \frac{h_0}{I_{zz}^\text{fid}}\left(\frac{c^4 d}{16\pi^2G f_\text{rot}^2}\right) 
\approx \scinum{2.36}{-6}\left(\frac{h_0}{10^{-25}}\right) \left(\frac{d}{1\,\text{kpc}}\right)\left( \frac{100\,\text{Hz}}{f_\text{rot}} \right)^2 \left(\frac{10^{38}\,\text{kg}\,\text{m}^2}{I_{zz}^\text{fid}}\right).
\end{equation}

If emission of gravitational radiation via the $l=m=2$ mass quadrupole is considered to be the sole
energy loss mechanism for a pulsar, then by equating the \gw luminosity \citep[see, e.g.,
Equation~(4) of][]{2014ApJ...785..119A}
\begin{equation}\label{eq:luminosity}
\dot{E}_\text{gw} = \frac{8\pi^2c^3}{5G} f_\text{rot}^2 h_0^2 d^2 \approx \scinum{6.07}{29} \left(\frac{f_\text{rot}}{100\,\text{Hz}}\right)^2 \left( \frac{h_0}{10^{-25}} \right)^2 \left(\frac{d}{1\,\text{kpc}} \right)^2\,\text{W},
\end{equation}
with the loss of kinetic energy inferred from the the first frequency derivative
$\dot{f}_{\text{rot}}$ of the pulsar
\begin{equation}
\dot{E}_\text{KE} = 4\pi^2 I_{zz}^\text{fid} f_\text{rot} |\dot{f}_{\text{rot}}| \approx \scinum{3.95}{30} \left(\frac{I_{zz}^\text{fid}}{10^{38}\,\text{kg}\,\text{m}^2}\right)\left(\frac{f_{\text{rot}}}{100\,\text{Hz}} \right)\left(\frac{|\dot{f}_{\text{rot}}|}{10^{-11}\,\text{Hz}\,\text{s}^{-1}} \right)\,\text{W},
\end{equation}
one can define the spin-down limit on $h_0$, where
\begin{equation}\label{eq:h0sd}
h_0^{\text{sd}} = \frac{1}{d}\left(\frac{5}{2} \frac{G I_{zz}^\text{fid}}{c^3}\frac{|\dot{f}_{\text{rot}}|}{f_{\text{rot}}}\right)^{1/2}
\approx \scinum{2.55}{-25} \left(\frac{1\,\text{kpc}}{d}\right)\left(\frac{I_{zz}^\text{fid}}{10^{38}\,\text{kg}\,\text{m}^2}\right)^{1/2}\left(\frac{100\,\text{Hz}}{f_\text{rot}}\right)^{1/2}\left(\frac{|\dot{f}_{\text{rot}}|}{10^{-11}\,\text{Hz}\,\text{s}^{-1}} \right)^{1/2}.
\end{equation}
By equating Equations~(\ref{eq:h0}) and (\ref{eq:h0sd}), we can rearrange and get spin-down limits
on $Q_{22}$ as
\begin{equation}\label{eq:q22sd}
Q_{22}^{\text{sd}} = \left(\frac{75}{4096\pi^5}\frac{I_{zz}^\text{fid}c^5}{G} \frac{\dot{f}_\text{rot}}{f^5_\text{rot}}\right)^{1/2} 
 \approx \scinum{4.66}{32} \left( \frac{I_{zz}^\text{fid}}{10^{38}\,\text{kg}\,\text{m}^2} \right)^{1/2} \left(\frac{100\,\text{Hz}}{f_\text{rot}}\right)^{5/2} \left(\frac{|\dot{f}_{\text{rot}}|}{10^{-11}\,\text{Hz}\,\text{s}^{-1}} \right)^{1/2}\,\text{kg}\,\text{m}^2,
\end{equation}
and on $\varepsilon$ as
\begin{equation}\label{eq:ellsd}
\varepsilon^{\text{sd}} = \left(\frac{5}{512\pi^4}\frac{c^5}{I_{zz}^\text{fid}G} \frac{\dot{f}_\text{rot}}{f^5_\text{rot}}\right)^{1/2} 
 \approx \scinum{6.03}{-6} \left( \frac{10^{38}\,\text{kg}\,\text{m}^2}{I_{zz}^\text{fid}} \right)^{1/2} \left(\frac{100\,\text{Hz}}{f_\text{rot}}\right)^{5/2} \left(\frac{|\dot{f}_{\text{rot}}|}{10^{-11}\,\text{Hz}\,\text{s}^{-1}} \right)^{1/2},
\end{equation}
where it is interesting to note that these are independent of the distance to the pulsar.

For a triaxial source not rotating about a principal axis, and emitting via both the $l=2$, $m=1$
and the $l=m=2$ quadrupole modes, the relations between the waveform amplitudes and phases given in
Equations~(\ref{eq:l2m1}) and (\ref{eq:l2m2}) and the source moment-of-inertia tensor components and
Euler orientation angle $\theta$ are described in Section~3.1 of \citet{2015MNRAS.453...53J}. We
will not repeat the relationships here, but note that how to convert between the two definitions is
described in detail in the Appendix of \citet{2015MNRAS.453.4399P}.

\section{Priors}\label{ap:priors}

In this appendix we will detail the prior probability distributions used on parameters by the {\it
Bayesian} and $5n$-{\it vector} analysis methods. The use of these priors for the {\it Bayesian}
search is discussed in \citet{2017arXiv170508978P}, and the motivation behind some of the prior
limits used are discussed in \citet{2015MNRAS.453...53J} and \citet{2015MNRAS.453.4399P}. For the
$5n$-{\it vector} pipeline, priors are set on signal initial phase $\phi_0$ and polarization
parameters $\psi$, $\cos\iota$, in the computation of upper limits.

For the gravitational-wave-specific orientation parameters for searches purely from the $l=m=2$
mode, the following priors have been used.\footnote{In the notation used here $\sim$ stands for
``has the probability distribution of,'' and $\mathcal{U}(a,b)$ is a continuous uniform distribution
with a constant probability $1/(b-a)$ for $x \in [a,b]$.} The initial rotational phase of the pulsar
at a given epoch $\phi_0$, the polarization angle $\psi$, and the cosine of the inclination angle
$\cos\iota$ have uniform priors\footnote{The polarization angle $\psi$, and orientation angle
$\iota$, have a joint prior that is uniform over a sphere, with degeneracies when thinking purely in
terms of the \gw waveforms described in \citet{2015MNRAS.453...53J}, but these can be
reparametrized to independent uniform priors if in terms of $\cos\iota$.} given by
\begin{align*}
\phi_0 &\sim \mathcal{U}(0, \pi), \\
\psi &\sim \mathcal{U}(0, \pi/2), \\
\cos\iota &\sim \mathcal{U}(-1,1).
\end{align*}
For the {\it Bayesian} search, the prior on the \gw amplitude $h_0$ is based on observed upper
limits, or sensitivity estimates, from previous LIGO and Virgo runs. The form of the prior is given
by a Fermi-Dirac-type probability distribution \citep[see, e.g., that used
in][]{2016MNRAS.455L..72M} as described in \citet{2017arXiv170508978P}, which has a flat region
followed by an exponential decay region but is nonzero for all positive values. It is defined as
\begin{equation}
        p(x|\sigma, \mu, I) = \begin{cases}\frac{1}{\sigma\ln{\left(1+e^{\mu/\sigma} \right)}}\left(e^{(x-\mu)/\sigma} + 1\right)^{-1} & \text{if } x \geqslant 0, \\
                0 & \text{otherwise},
               \end{cases}
\end{equation}
where $\mu$ gives the value at which the distribution decays to 50\% of its maximum value and
$\sigma$ controls the width of the band over which the bulk of the decay happens. The band around
$\mu$ over which the probability density falls from 97.5\% to 2.5\% of its peak value is given by
$\mu \pm 7.33\mu/2r$, where $r=\mu/\sigma$. In our case we specify that this fall-off happens over a
range that is 40\% of the value of $\mu$, so that $r=7.33/(2 \times 0.4) = 9.1625$. The value of
$\mu$ is set by finding the value that produces a specific bound within which 95\% of the
probability is constrained (bounded by zero at the lower end) given the previous value of $r$. The
specific bound is that based on the sensitivity for each pulsar (i.e., the 95\% upper limits on
$h_0$, see Appendix~\ref{ap:sensitivity}) that would have been expected if using data from the sixth
LIGO science run and fourth Virgo science run, scaled up by a factor of 25 to be conservative and
make sure that the likelihood is well within the flat part of the prior distribution, while
disfavoring arbitrarily large values.\footnote{A discussion about a choice between a uniform prior
and a uniform in logarithm prior for the amplitude parameter is given in Appendix~B of
\citet{2017PhRvD..96d2001I}.}

For the searches that include both the $l=2$, $m=1,\,2$ modes the phase and orientation angle priors
have been given by
\begin{align*}
 \Phi_{21}^C &\sim \mathcal{U}(0, 2\pi), \\
 \Phi_{22}^C &\sim \mathcal{U}(0, 2\pi), \\
 \psi &\sim \mathcal{U}(0, \pi/2), \\
 \cos\iota &\sim \mathcal{U}(-1,1).
\end{align*}
As discussed above, in the {\it Bayesian} method the priors on the amplitude parameters $C_{21}$ and
$C_{22}$ have used Fermi-Dirac probability distributions for which the parameters have been set in
the same way as done for $h_0$. However, in this case the sensitivity estimate used for $h_0$ is
assumed to be valid for $C_{21}$ and $C_{22}$, while in reality there are factors of a few
differences. These differences are allowable given the scaling factor used and the sensitivity
improvements over S6.

In our searches we make use of the pulsar rotational phase parameters (frequency, frequency
derivatives, sky location, proper motion, and Keplerian and relativistic binary system orbital
parameters if relevant) derived from electromagnetic observation of pulse times of arrival. These
parameters are obtained by fitting the phase model to the times of arrival using software such as
{\sc Tempo2} \citet{2006MNRAS.369..655H} to produce ephemeris files, and these fits include
uncertainty estimates. In most cases, and where it is computationally feasible, for any combination
of parameters in the ephemeris files that have been refit (i.e., a new estimate has been performed
using data that matched the requirements of our search, such as being concurrent with the LIGO
observing runs) we include a multivariate Gaussian prior in our analysis, for which the diagonal of
the covariance matrix is derived from the uncertainties in the ephemeris file and taking them to be
one standard deviation values. In the prior covariance matrix we assume no correlations between
parameters except in two pairs of cases for pulsars in binary systems; for very low eccentricity
systems ($e < 0.001$) with refitted uncertainties on both the time and angle of periastron, or with
refitted values on the period and time derivative of the angle of periastron, the covariance matrix
is set such as to make these pairs fully correlated.

As described in \citet{2010ApJ...713..671A} and \citet{2014ApJ...785..119A,2017ApJ...839...12A},
there are some pulsars for which we can place tighter constraints on their orientation. In
particular, the inclination angle and \gw polarization angle can be assumed to be measured by
modeling X-ray observations of their surrounding pulsar wind nebulae \citep{Ng:2004,Ng:2008}. In
this analysis, for PSR J0205+6449, PSR J0534+2200, PSR J0835\textminus4510, PSR J1952+3252, and PSR
J2229+6114, in addition to a search using the above priors, we also perform parameter estimation
using the restricted priors given in Table~3 of \citet{2017ApJ...839...12A}, based on values taken
from \citet{Ng:2008}. In these cases the priors are on the inclination angle $\iota$ rather than its
cosine. The prior probability distribution on $\psi$ is a unimodal Gaussian, but that on $\iota$ is
given by the sum of a pair of Gaussian distributions with different means, which is required to
account for the fact that rotation directions of the stars are unknown \citep{2015MNRAS.453...53J}.

\section{Sensitivity estimates}\label{ap:sensitivity}

Here we will describe the expected sensitivity of the {\it Bayesian} analysis in searches for
signals purely from the $l=m=2$ mode, and for coherent searches for signals at both the $l=2$,
$m=1,\,2$ modes. We define the expected sensitivity based on the observation time ($T_{\text{obs}}$)
weighted noise power spectral density $S_n(f)$ as a function of frequency $f$, such that for a
single detector
\begin{equation}\label{eq:hsens}
\langle h(f) \rangle = D \sqrt{\frac{S_n(f)}{T_{\text{obs}}}},
\end{equation}
where in our case $\langle h(f) \rangle$ is the expected 95\% credible upper limit on amplitude and
$D$ is an empirically derived scaling factor \citep[similar to the sensitivity depth defined
in][]{2015PhRvD..91f4007B}. When combining data from multiple detectors and observing runs, for
which the power spectral densities will be different, we take the harmonic mean of the time-weighted
power spectral densities. For example, for a set of different noise power spectral densities
$S_{n_i}(f)$ associated with observation times $T_{{\text{obs}}_i}$ we would have
\begin{equation}
\langle h(f) \rangle = D \left( \sum_{i=1}^N \left[\frac{S_{n_i}(f)}{T_{{\text{obs}}_i}} \right]^{-1} \right)^{-1/2}.
\end{equation}

\begin{figure*}
\epsscale{1.15}
\plotone{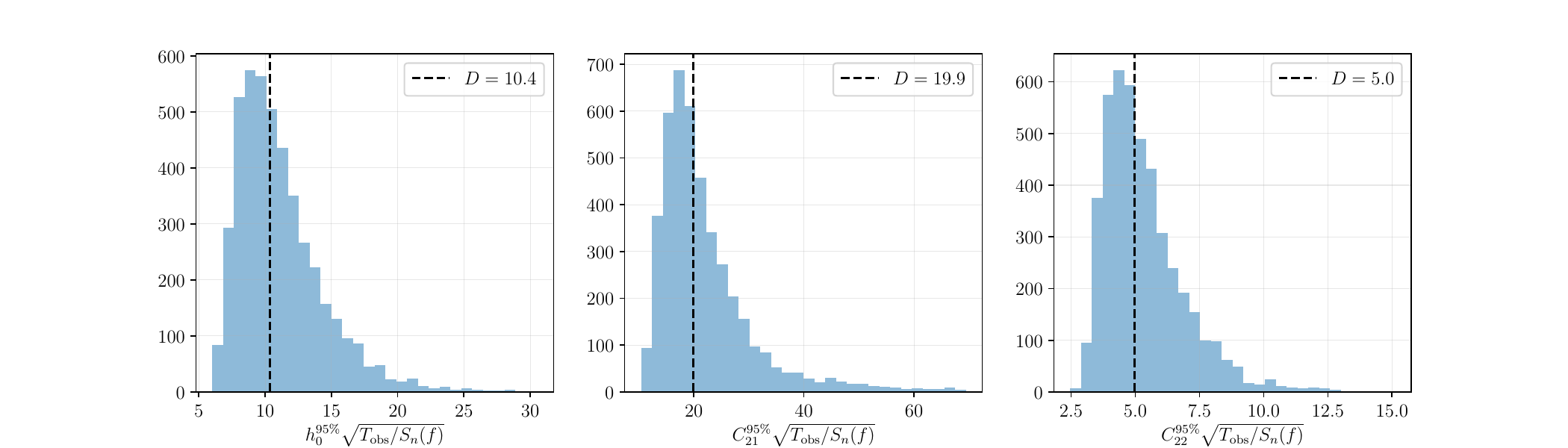}
\caption{Distributions of 95\% credible upper limits on $h_0$ (left), $C_{21}$ (middle), and $C_{22}$
(right) scaled by the observation times and noise power spectral density for a set of simulations
consisting of Gaussian noise. To average over effects of different antenna patterns in performing
parameter estimation, each simulation assumes a random source sky location for a uniform distribution
over the sky.
\label{fig:senshists}}
\end{figure*}

For a search for emission from the $l=m=2$ mode, where the limit is on the \gw amplitude $h_0$ (see
Equation~(\ref{eq:h0})), it was shown in \citet{Dupuis:2005} that $D \approx 10.8 \pm 0.2$, based on
the simulations containing purely Gaussian noise with variance drawn from a known power spectral
density, marginalized over orientations and averaged over the sky. If we instead take the median
rather than the mean over a similar set of simulations, to suppress any outlier values, we find $D
\approx \HZEROSENS$ (see left panel of Figure~\ref{fig:senshists}), which is used here in
producing the sensitivity curve in Figure~\ref{fig:resultsh0}.

\begin{figure}
\epsscale{0.48}
\plotone{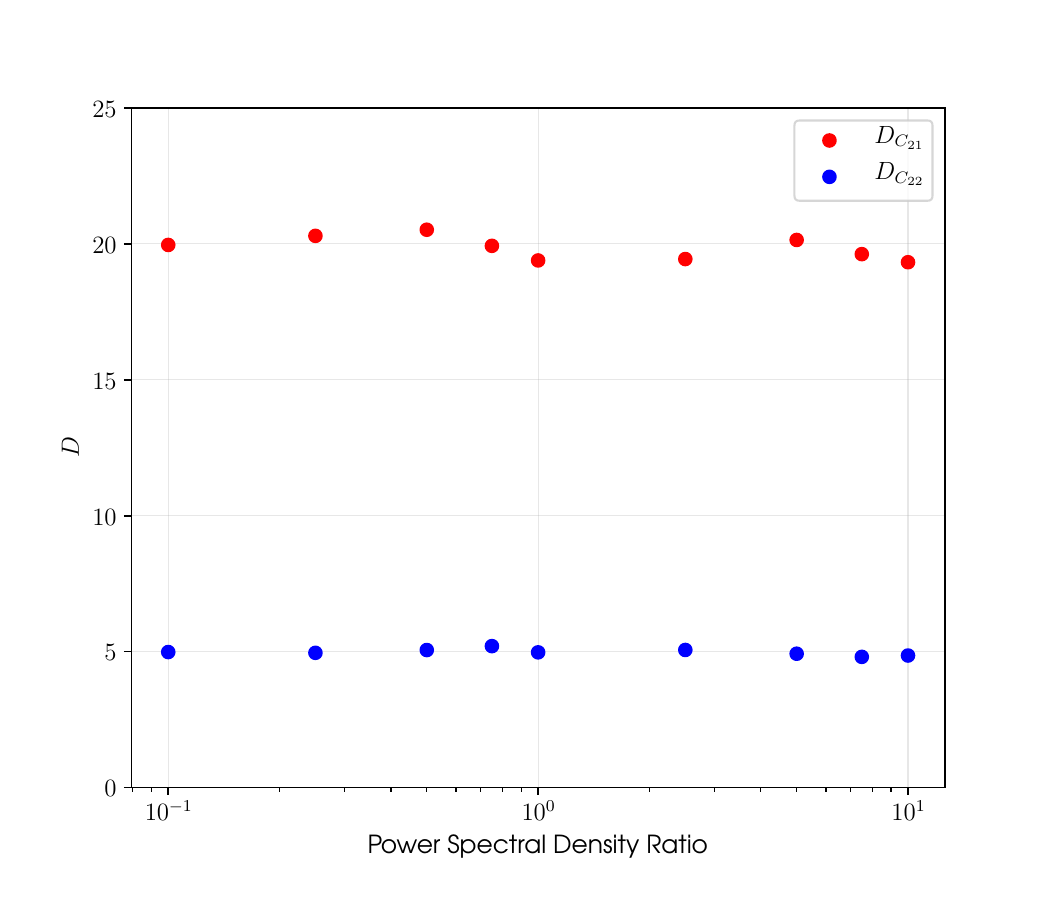}
\caption{The $D$ scale factor for the $C_{21}$ and $C_{22}$ upper limits as a function of the power
spectral density ratio between the data at equivalents of the rotation frequency and twice the
rotation frequency. \label{fig:sensratio}}
\end{figure}

To estimate the sensitivity to the $C_{21}$ and $C_{22}$ amplitude parameters for an $l=2$,
$m=1,\,2$ mode search, we have performed similar simulations to those described above. A search
including both modes is not completely independent for each mode, as there are common orientation
parameters. Hence, we also wanted to investigate whether the sensitivity at either amplitude is
affected by the noise level at the other amplitude. We generated simulations consisting of
independent Gaussian noise in two data streams: one equivalent to the data at the rotation frequency
and another equivalent to the data at twice the rotation frequency. For the data stream at twice the
rotation frequency the noise was always drawn from a Gaussian distribution with the same variance
defined by a power spectral density of $10^{-48}\,\text{Hz}^{-1/2}$. For the data stream at the
rotation frequency we created multiple sets of 500 instantiations where the noise was drawn from a
Gaussian distribution with a variance defined by a power spectral density of
$10^{-48}x\,\text{Hz}^{-1/2}$, where for each set of 500 $x$ was a different factor between 0.1 and
10. The $D$ scale factor from Equation~(\ref{eq:hsens}) for both the $C_{21}$ and $C_{22}$ amplitude
upper limit for each set of 500 simulations and as a function of $x$ is shown in
Figure~\ref{fig:sensratio}. It can be seen that there is no obvious correlation between the power
spectral density ratio $x$ and the value of $D$, which suggests that the upper limits on the two
amplitudes are actually largely independent.

We see from Figures~\ref{fig:senshists} and \ref{fig:sensratio} that the value of $D$ used to
estimate the sensitivity for $C_{21}$ is \CTOSENS, and the value of $D$ used to estimate the
sensitivity for $C_{22}$ is \CTTSENS. These values have been used when producing the sensitivity
curves in Figure~\ref{fig:resultsC}.

\section{Mixed Bayesian/Frequentist upper limit computation for the $5n$-vector method}\label{ap:5vecul}

Given a measured value ${S}^*$ of a detection statistic $\cal{S}$, the frequentist upper limit at a
given confidence level $\alpha$ is defined as that value of signal amplitude $h_\text{ul}$ such that
a signal with amplitude $h_0>h_\text{ul}$ produces a value of the detection statistic bigger than
${S}^*$ in a fraction $\alpha$ of a large number of repeated experiments: $P({S}>{S}^*\vert
h_{0}>h_\text{ul})=\alpha$. Typically, the upper limit is computed using Neyman's rule for the
construction of confidence intervals \citep{Neyman:1938}. This classical frequentist upper limit has
the following well-known and unpleasant feature: if the value of the detection statistic ${S}^*$
falls in the first 1-$\alpha$ quantile of its noise-only distribution, the resulting upper limit is
exactly zero. This behavior, although legitimate in the frequentist framework, poses a problem, for
instance, when upper limits obtained in the analysis of datasets with different sensitivity are
compared. It may happen that, due to a noise fluctuation, the upper limit set for the more noisy
data is below that computed for the less noisy one. This kind of problem may happen also for
Bayesian upper limits, but it is exacerbated in the classical frequentist case.   

The unwanted features of the classical Neyman's construction have been overcome in the
Feldman--Cousins unified approach, where, using the freedom inherent in Neyman's construction, a
method to obtain a unified set of classical confidence intervals for computing both upper limits and
two-sided confidence intervals has been obtained \citep{1998PhRvD..57.3873F}. The Feldman--Cousins
approach sometimes is difficult to implement and, similarly to the Neyman's approach, does not
allow accounting for nonuniform prior distributions for nuisance parameters.

We have developed an alternative method for setting upper limits on signal amplitude that keeps the
advantages of the frequentist approach, like the ease of implementation and computational speed,
while avoiding its problems. The basic idea is that of computing the posterior distribution of the
signal amplitude conditioned to the measured value of the detection statistic. The main steps of the
procedure can be summarized as follows. 

\begin{figure}
\epsscale{0.80}
\plotone{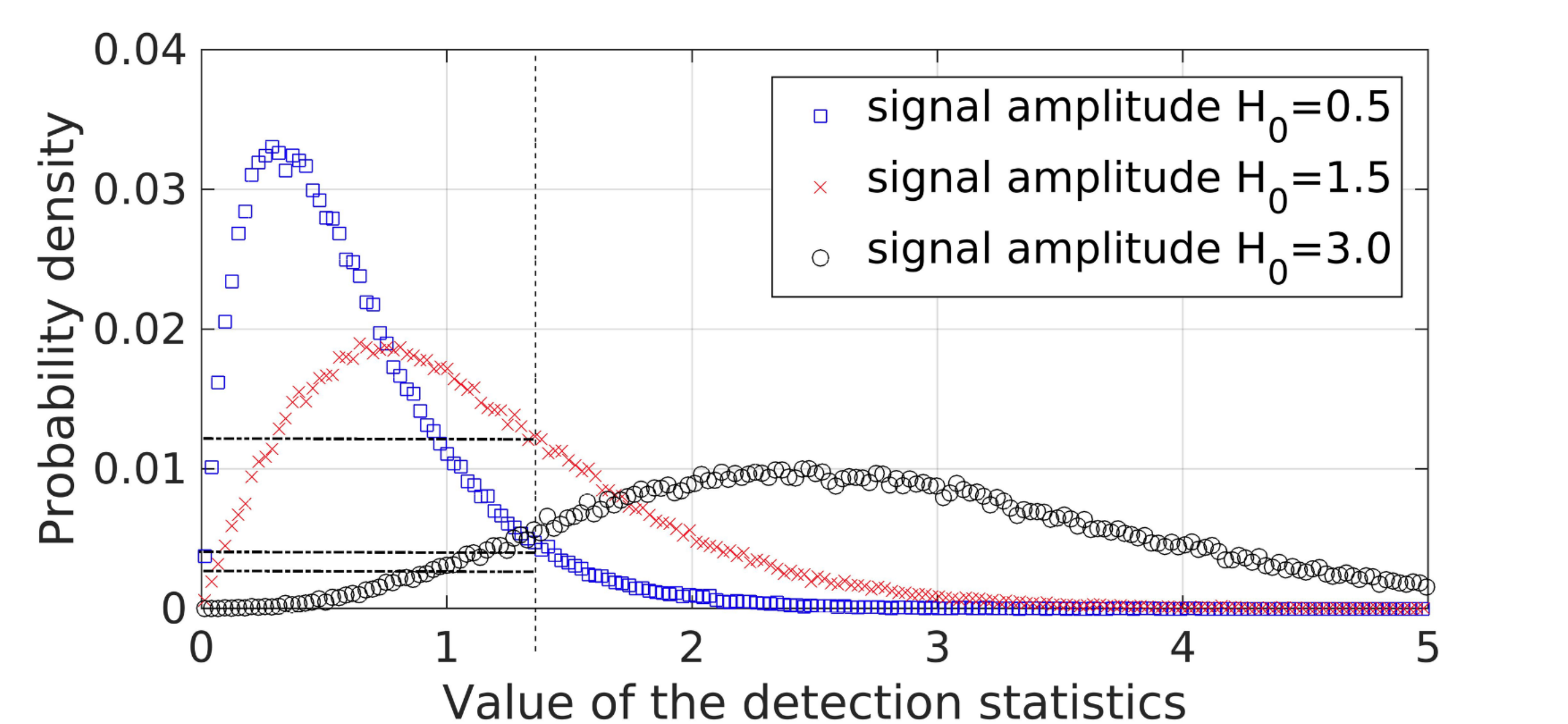}
\caption{Probability distributions of the detection statistic $\cal{S}$ after having injected into Gaussian noise with $\sigma$=1 signals with three different amplitudes. Given the measured value of the detection statistic $\cal{S}^*$ (shown by the vertical dashed line), the corresponding  values of probability density for the various signal amplitudes are determined (shown by the horizontal dot-dashed lines).
 \label{fig:ul1}}
\end{figure}

We consider a set of possible signal amplitudes $H_0$. For each amplitude we generate several
signals with polarization parameters distributed according to given prior distributions, and for each
signal we compute the corresponding value of the detection statistic. Hence, the probability
distribution of the detection statistic, for the different signal amplitudes, can be built; see
Figure~\ref{fig:ul1}.
For each distribution we determine the value corresponding to the measured detection statistic
$p({S}^* \vert H_0)$. By multiplying each value by the prior probability density of the signal
amplitude, $p(H_0)$, and normalizing, we obtain the posterior probability distribution for the
signal amplitude: $p(H_0 \vert {S}^*)\propto p({S}^*\vert H_0) p(H_0)$, see Figure~\ref{fig:ul2}.
\begin{figure}
\epsscale{0.48}
\plotone{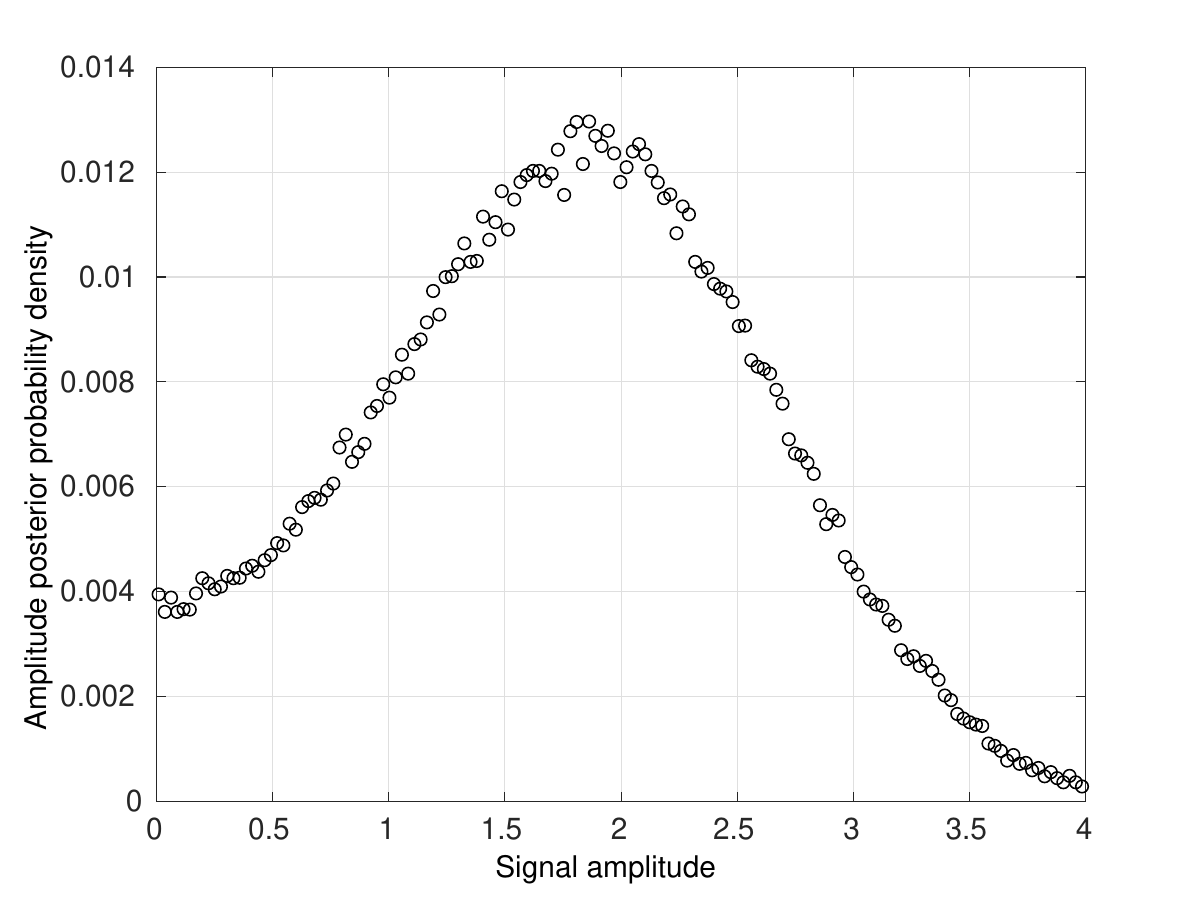}
\caption{Posterior probability distribution of the signal amplitude for the given measured value \cal{S}$^*$ of the detection statistic.}
\label{fig:ul2}
\end{figure}

\begin{figure}
\epsscale{0.48}
\plotone{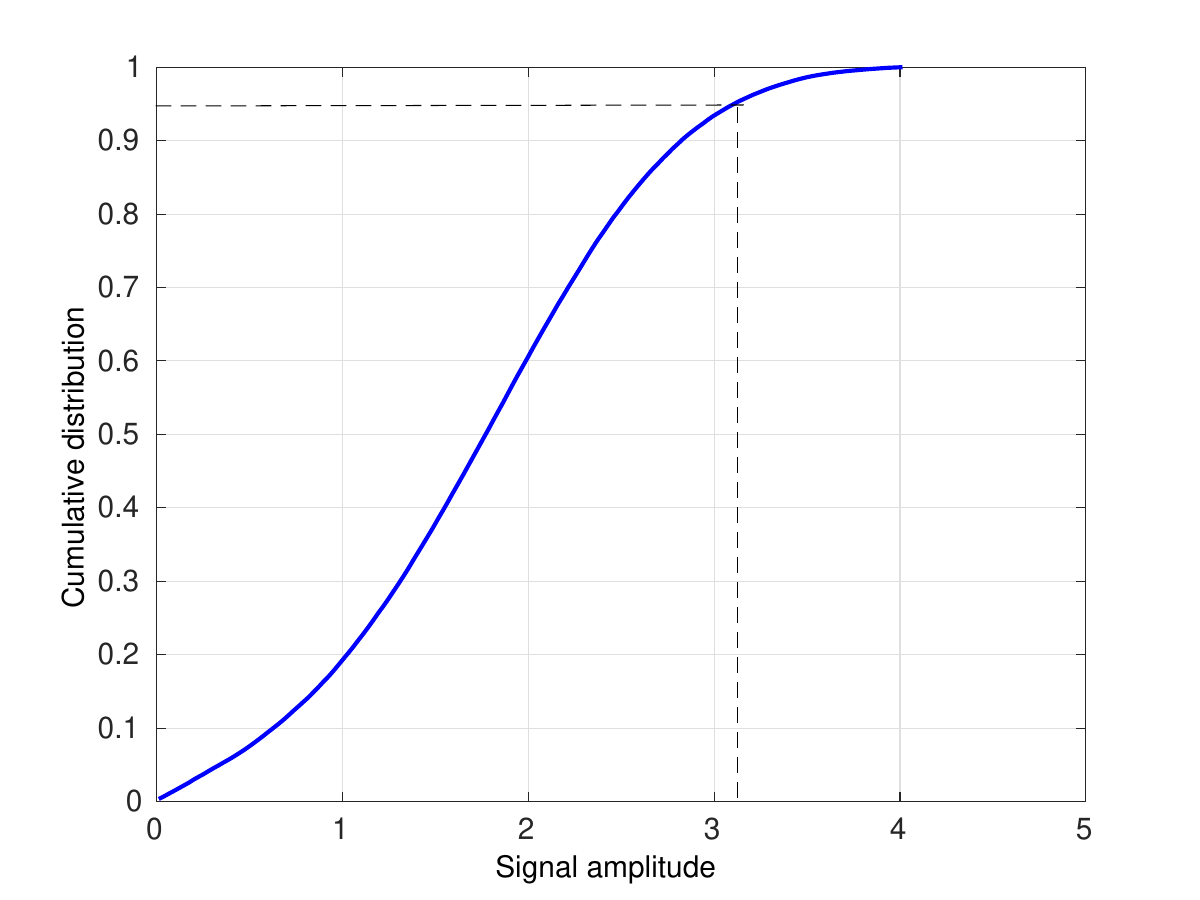}
\caption{Cumulative posterior probability distribution of the signal amplitude. The amplitude value corresponding to 95\% of the cumulative is the wanted credible upper limit.}
 \label{fig:ul3}
\end{figure}
We then calculate the cumulative probability distribution and obtain the amplitude value corresponding to a given probability, e.g., 0.95; see Figure~\ref{fig:ul3}. This is the 95\% credible upper limit. 

\section{Amplitude conversion factors for the $5n$-vector method}\label{ap:5vecampl}

The $5n$-vector method uses a nonstandard formalism to describe the \gw signal, based on the
concept of polarization ellipse \citep{2010CQGra..27s4016A,2011ApJ...737...93A,2014ApJ...785..119A}.
In this formalism the signal strain is given by the real part of
\begin{equation}\label{eq:hoft_pol}
h(t)=H_0(H_+{\bf e}^{+}+H_{\times}{\bf e}^{\times})e^{\imath \left(\omega_0(t)t+\Phi_0\right)}  
\end{equation}
where $\omega_0(t)$ is the signal angular frequency, ${\bf e}^{+/\times}$ are the two basis polarization
tensors, $\Phi_0$ is the signal phase at the time $t=0$, and the two complex amplitudes
$H_+,~H_{\times}$ are given by
\begin{equation}
H_+=\frac{\cos{2\psi}-\imath \eta \sin{2\psi}}{\sqrt{1+\eta^2}},~H_{\times}=\frac{\sin{2\psi}+\imath \eta \cos{2\psi}}{\sqrt{1+\eta^2}},
\label{eq:Hc}
\end{equation}
in which $\eta \in [-1,1]$ is the ratio of the polarization ellipse semi-minor to semi-major axis
and the polarization angle $\psi$ defines, as usual, the direction of the major axis with respect to
the celestial parallel of the source (measured counterclockwise). The signal described by
Equation~(\ref{eq:hoft_pol}) is general, i.e., does not assume any specific emission mechanism by a
spinning neutron star. Assuming a triaxial star spinning about a principal axis of inertia, the
overall amplitude $H_0$ is related to the standard $h_0$ by
\begin{equation}\label{eq:H0h0}
h_0=\frac{2H_0}{\sqrt{1+6 \cos^2 \iota+ \cos^4 \iota}}.
\end{equation}
For the emission at the star's rotational frequency of the $l=2,~m=1$ harmonic mode (see Equation
(\ref{eq:l2m1})), the relation between $H_0$ and the amplitude $C_{21}$ is given by
\begin{equation}\label{eq:H0C21}
C_{21}=\frac{2H_0}{\sqrt{1-\cos^4{\iota}}}
\end{equation}
As discussed in, e.g., \citet{2014ApJ...785..119A}, upper limits are computed on $H_0$ and then
converted to $h_0$ or $C_{21}$ using Equations~(\ref{eq:H0h0}) and (\ref{eq:H0C21}), where the
functions of $\iota$ are replaced by their mean value: $h^{95\%}_0\simeq 1.37H^{95\%}_0$, and
$C^{95\%}_{21}\simeq 1.31H^{95\%}_0$.

\bibliographystyle{aasjournal}
\bibliography{TargetedPulsarPaper}

\section*{Erratum: ``Searches for Gravitational Waves from Known Pulsars at Two Harmonics in 2015-2017 LIGO Data'' \lowercase{(\href{https://doi.org/10.3847/1538-4357/aa677f}{2019, \uppercase{A}p\uppercase{J}, 879, 1, 10})}}

\begin{figure*}
\epsscale{1.0}
\plotone{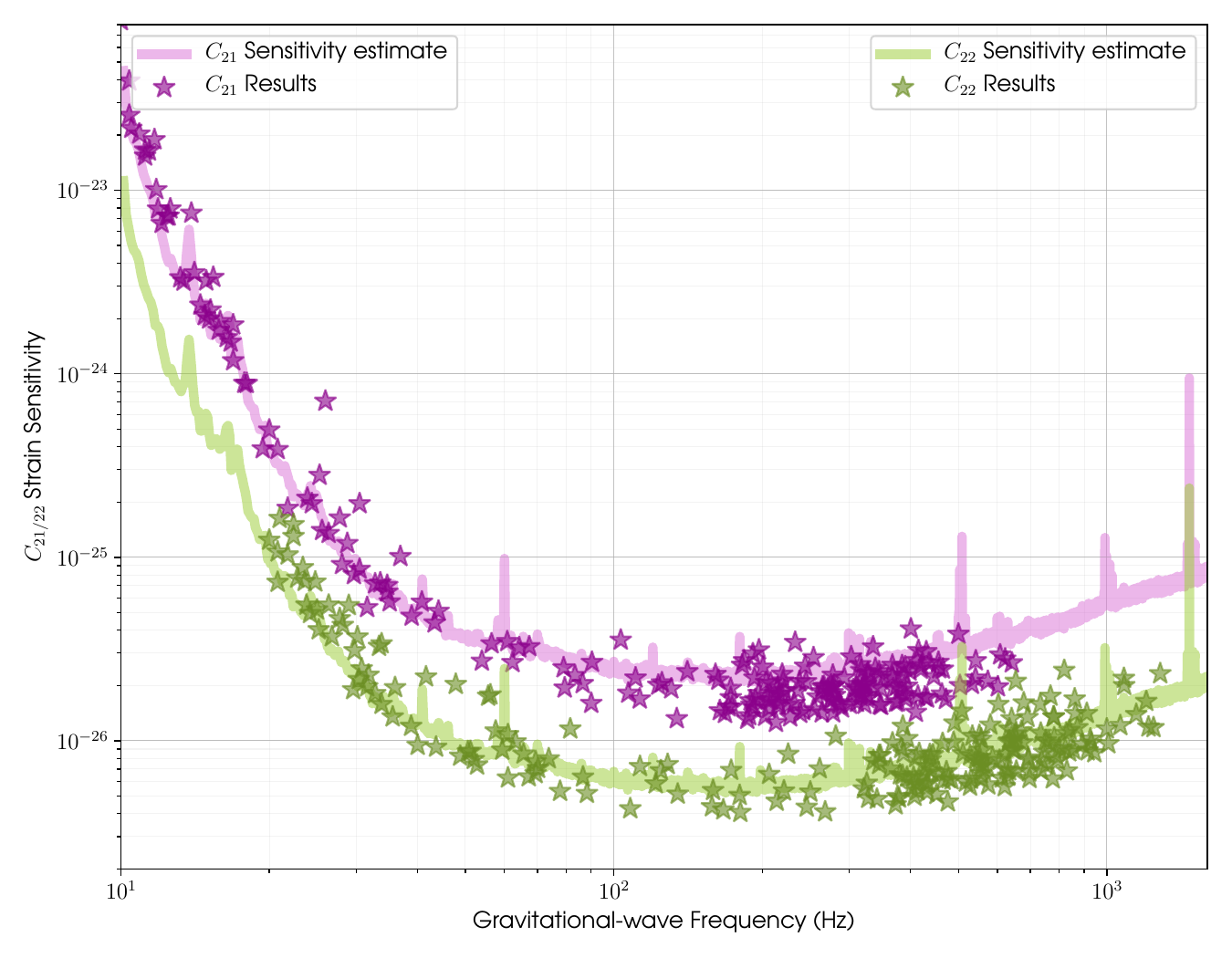}
\caption{Upper limits on $C_{21}$ and $C_{22}$ for 221 pulsars. The stars show the observed 95\% credible upper limits on observed amplitudes for each pulsar. The solid lines show an estimate of the expected sensitivity of the searches.\label{fig:ul12}}
\end{figure*}

Two analysis errors have been identified that affect the results for a handful of the high-value pulsars given in Table~1 of \citet{Abbott:2019}. One affects the {\it Bayesian} analysis for the five pulsars that glitched during the analysis period, and the other affects the $5n$-{\it vector} analysis for J0711\textminus6830.
Updated results after correcting the errors are shown in Table~\ref{tab:highvalue}, which now supersedes the results given for those pulsars in
Table~1 of \citet{Abbott:2019}. Updated versions of figures can be seen in Figure~\ref{fig:ul12}, \ref{fig:ul2}, \ref{fig:sdrat} and \ref{fig:ell}.

\paragraph{Bayesian analysis}
For the glitching pulsars, the signal phase evolution caused by the glitch was wrongly applied twice and was therefore not
consistent with our expected model of the pulsar phase. This error did not affect the $\mathcal{F}/\mathcal{G}$-statistic or
$5n$-{\it vector} analysis.

Analyses of the five pulsars PSR J0205+6449, J0534+2200,
J0835\textminus4510, J1028\textminus5819, and J1718\textminus3825 have been repeated after correcting for the error. There are small quantitative differences in the results, but the changes do not affect the main conclusions of the paper. The largest differences are for PSR J0835\textminus4510 (the Vela pulsar), for which the updated upper limits from the {\it Bayesian} method are found to be between 1.1 to 2 times larger than those obtained when the error was present. This appears primarily to be due the error leading to the decohering of a strong spectral line in the LIGO Livingston detector and thus lowering the amplitude limit.

\paragraph{$5n$-{\it vector} analysis}
An error was also identified in the settings of the $5n$-{\it vector} analysis, which affected the upper limit computation at the rotation frequency for $C^{95\%}_{21}$ of J0711\textminus6830. Specifically, we found an incorrect choice for the range of amplitudes used to inject simulated signals in the O2 data. The updated upper limit is about 2.5 times worse than that obtained when the error was present. This error did not affect the {\it Bayesian} or $\mathcal{F}/\mathcal{G}$-{\it statistic} results.

\clearpage
\startlongtable
\movetabledown=1.9cm
\begin{longrotatetable}
\begin{deluxetable}{lrllllllllllll}
\tablecaption{Limits on Gravitational-wave Amplitude, and Other Derived Quantities, for 5 High-value Pulsars from the Three Analysis Methods.\label{tab:highvalue}}
\tabletypesize{\footnotesize}
\tablehead{
\colhead{Pulsar Name} & 
\colhead{$f_{\text{rot}}$} & 
\colhead{$\dot{P}_{\text{rot}}$} & 
\colhead{Distance} & 
\colhead{$h_0^{\text{sd}}$} & 
\colhead{Analysis} & 
\colhead{$C_{21}^{95\%}$} & 
\colhead{$C_{22}^{95\%}$} & 
\colhead{$h_0^{95\%}$} & 
\colhead{$Q_{22}^{95\%}$} & 
\colhead{$\varepsilon^{95\%}$} & 
\colhead{$h_0^{95\%}/h_0^{\text{sd}}$} & 
\colhead{Statistic\tablenotemark{a}} & 
\colhead{Statistic\tablenotemark{b}}\\ 
\colhead{(J2000)} & 
\colhead{(Hz)} & 
\colhead{(s\,s$^{-1})$} & 
\colhead{(kpc)} & 
\colhead{~} & 
\colhead{Method} & 
\colhead{~} & 
\colhead{~} & 
\colhead{~} & 
\colhead{(kg\,m$^2$)} & 
\colhead{~} & 
\colhead{~} & 
\colhead{$^{l=2, m=1,2}$} & 
\colhead{$^{l=2, m=2}$}}
\startdata
\multirow{3}{*}{J0205+6449\tablenotemark{c}} & \multirow{3}{*}{15.2} & \multirow{3}{*}{\scinum{1.9}{-13}} & \multirow{3}{*}{2.00 (c)} & \multirow{3}{*}{\scinum{6.9}{-25}} & Bayesian & \scinum{2.2(1.6)}{-24} & \scinum{2.2(2.9)}{-26} & \scinum{4.5(5.7)}{-26} & \scinum{7.2(9.0)}{33} & \scinum{0.9(1.2)}{-4} & 0.065(0.082) & -4.8(-4.7) & -2.8(-2.6) \\  &  &  &  &  & $\mathcal{F}$-statistic & \scinum{2.2}{-24} & \scinum{4.5}{-26} & \scinum{8.8}{-26} & \scinum{1.4}{34} & \scinum{1.8}{-4} & 0.13 & 0.71 & 0.26 \\  &  &  &  &  & 5$n$-vector & \nodata & \nodata & \scinum{2.9(4.5)}{-26} & \scinum{4.6(7.1)}{33} & \scinum{5.9(9.2)}{-5} & 0.042(0.065) & \nodata & 0.41 \\ 
\hline 
\multirow{3}{*}{J0534+2200\tablenotemark{c}} & \multirow{3}{*}{29.7} & \multirow{3}{*}{\scinum{4.2}{-13}} & \multirow{3}{*}{2.00} & \multirow{3}{*}{\scinum{1.4}{-24}} & Bayesian & \scinum{8.1(5.9)}{-26} & \scinum{8.9(7.6)}{-27} & \scinum{1.9(1.5)}{-26} & \scinum{7.8(6.3)}{32} & \scinum{1.0(0.8)}{-5} & 0.013(0.011) & -5.2(-5.3) & -2.6(-2.6) \\  &  &  &  &  & $\mathcal{F}$-statistic & \scinum{1.6(1.1)}{-25} & \scinum{1.1(1.1)}{-26} & \scinum{2.2(1.3)}{-26} & \scinum{9.1(5.4)}{32} & \scinum{1.2(0.7)}{-5} & 0.015(0.0091) & 0.32(0.18) & 0.65(0.87) \\ 
 &  &  &  &  & 5$n$-vector & \scinum{1.7(1.3)}{-25} & \nodata & \scinum{2.9(2.9)}{-26} & \scinum{1.2(1.2)}{33} & \scinum{1.6(1.6)}{-5} & 0.02(0.02) & 0.70 & 0.45 \\ 
\hline
\multirow{3}{*}{J0711-6830\tablenotemark{c}} & \multirow{3}{*}{182.1} & \multirow{3}{*}{\scinum{1.4}{-20}} & \multirow{3}{*}{0.11 (b)} & \multirow{3}{*}{\scinum{1.2}{-26}} & Bayesian & \scinum{2.6}{-26} & \scinum{7.0}{-27} & \scinum{1.5}{-26} & \scinum{9.3}{29} & \scinum{1.2}{-8} & 1.3 & -3.1 & -1.9 \\  &  &  &  &  & $\mathcal{F}$-statistic & \nodata  & \nodata & \nodata & \nodata & \nodata & \nodata & \nodata & \nodata \\  &  &  &  &  & 5$n$-vector & \scinum{3.0}{-26} & \nodata & \scinum{1.5}{-26} & \scinum{9.1}{29} & \scinum{1.2}{-8} & 1.3 & 0.79 & 0.39 \\ 
\hline 
\multirow{3}{*}{J0835\textminus4510\tablenotemark{c}} & \multirow{3}{*}{11.2} & \multirow{3}{*}{\scinum{1.2}{-13}} & \multirow{3}{*}{0.29 (j)} & \multirow{3}{*}{\scinum{3.3}{-24}} & Bayesian & \scinum{1.5(1.4)}{-23} & \scinum{1.3(1.0)}{-25} & \scinum{2.4(2.1)}{-25} & \scinum{1.0(0.9)}{34} & \scinum{1.3(1.1)}{-4} & 0.073(0.062) & -3.3(-3.1) & -1.8(-2.1) \\ 
 &  &  &  &  & $\mathcal{F}$-statistic & \scinum{1.3(1.1)}{-23} & \scinum{1.1(0.9)}{-25} & \scinum{2.6(2.0)}{-25} & \scinum{1.1(0.8)}{34} & \scinum{1.4(1.1)}{-4} & 0.078(0.06) & 0.75(0.75) & 0.75(0.75) \\ 
 &  &  &  &  & 5$n$-vector & \nodata & \nodata & \scinum{2.3(2.4)}{-25} & \scinum{9.7(9.9)}{33} & \scinum{1.3(1.3)}{-4} & 0.07(0.071) & \nodata & 0.41 \\ 
\hline
\multirow{3}{*}{J1028\textminus5819} & \multirow{3}{*}{10.9} & \multirow{3}{*}{\scinum{1.6}{-14}} & \multirow{3}{*}{1.42 (b)} & \multirow{3}{*}{\scinum{2.4}{-25}} & Bayesian & \scinum{2.0}{-23} & \scinum{1.0}{-25} & \scinum{2.4}{-25} & \scinum{5.2}{34} & \scinum{6.7}{-4} & 1 & -3.8 & -2.2 \\ 
 &  &  &  &  & $\mathcal{F}$-statistic & \nodata & \nodata & \nodata & \nodata & \nodata & \nodata & \nodata & \nodata \\ 
 &  &  &  &  & 5$n$-vector & \nodata & \nodata & \scinum{1.9}{-25} & \scinum{4.1}{34} & \scinum{5.3}{-4} & 0.8 & \nodata & 0.40 \\ 
\hline
\multirow{3}{*}{J1718\textminus3825} & \multirow{3}{*}{13.4} & \multirow{3}{*}{\scinum{1.3}{-14}} & \multirow{3}{*}{3.49 (b)} & \multirow{3}{*}{\scinum{9.7}{-26}} & Bayesian & \scinum{3.2}{-24} & \scinum{3.7}{-26} & \scinum{7.8}{-26} & \scinum{2.8}{34} & \scinum{3.6}{-4} & 0.8 & -5.7 & -2.5 \\ 
 &  &  &  &  & $\mathcal{F}$-statistic & \nodata & \nodata & \nodata & \nodata & \nodata & \nodata & \nodata & \nodata \\ 
 &  &  &  &  & 5$n$-vector & \nodata & \nodata & \scinum{6.5}{-26} & \scinum{2.3}{34} & \scinum{3.0}{-4} & 0.67 & \nodata & 0.67 \\ 
\enddata
\tablecomments{For references and other notes see Table~2 in \citet{Abbott:2019}. Values in parentheses are those produced using the restricted orientation priors described in Section~2.2.4 of \citet{Abbott:2019}.}
\tablenotetext{a}{For the {\it Bayesian} method this column shows the base-10 logarithm of the Bayesian odds, $\mathcal{O}$, comparing a coherent signal model at both the $l=2$, $m=1,2$ modes to incoherent signal models. For the $\mathcal{F}$-/$\mathcal{G}$-{\it statistic} method this column shows the false alarm probability for a signal just at the $l=2$, $m=1$ mode, assuming that the $2\mathcal{F}$ value has a $\chi^2$ distribution with 4 degrees-of-freedom and the $2\mathcal{G}$ value has a $\chi^2$ distribution with 2 degrees-of-freedom. For the 5$n$-{\it vector} method this column shows the $p$-value for a search for a signal at just the $l=2$, $m=1$ mode, where the null hypothesis being tested is that the data is consistent with pure Gaussian noise.}
\tablenotetext{b}{This is the same as in footnote \tablenotemark{a}, but for all the methods the assumed signal model is from the $l=m=2$ mode.}
\tablenotetext{c}{The observed $\dot{P}$ has been corrected to account for the relative motion between the pulsar and observer.}
\end{deluxetable}
\end{longrotatetable}

\begin{figure*}
\epsscale{1.0}
\plotone{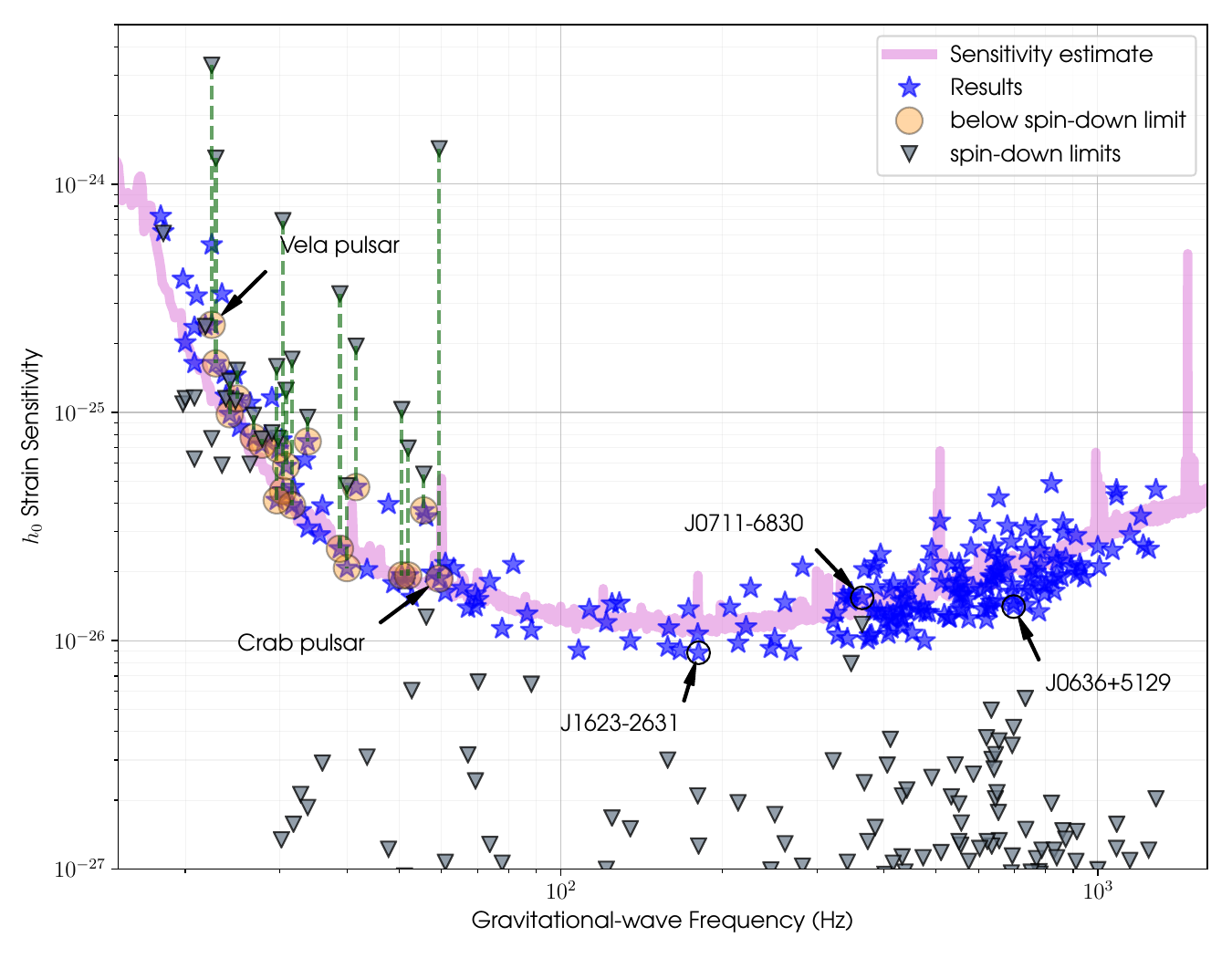}
\caption{Upper limits on $h_0$ for 221 pulsars. The stars show the observed 95\% credible upper limits on observed amplitude for each pulsar. The solid line shows an estimate of the expected sensitivity of the search. Triangles show the limits on gravitational-wave amplitude derived from each pulsar's observed spin-down.\label{fig:ul2}}
\end{figure*}

\begin{figure}
\epsscale{0.65}
\plotone{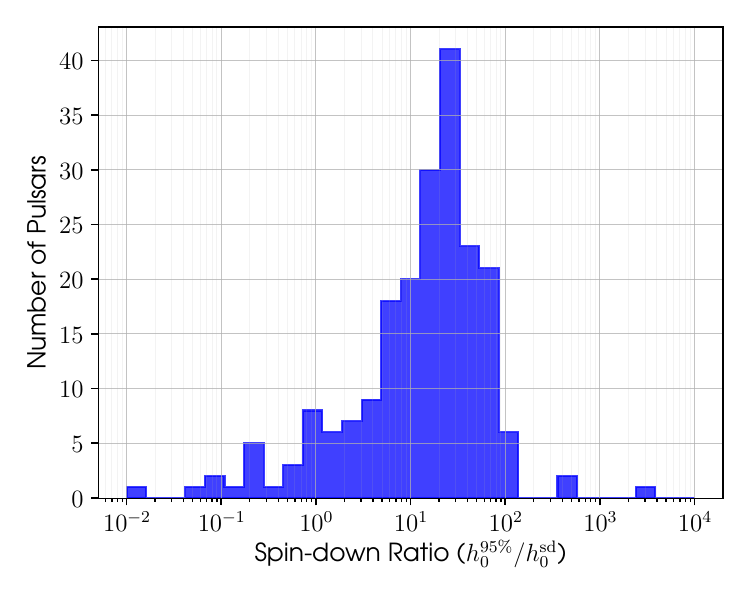}
\caption{Histogram of ratios of upper limits on $h_0$ compared to the spin-down limit.\label{fig:sdrat}}
\end{figure}

\begin{figure*}
\epsscale{1.0}
\plotone{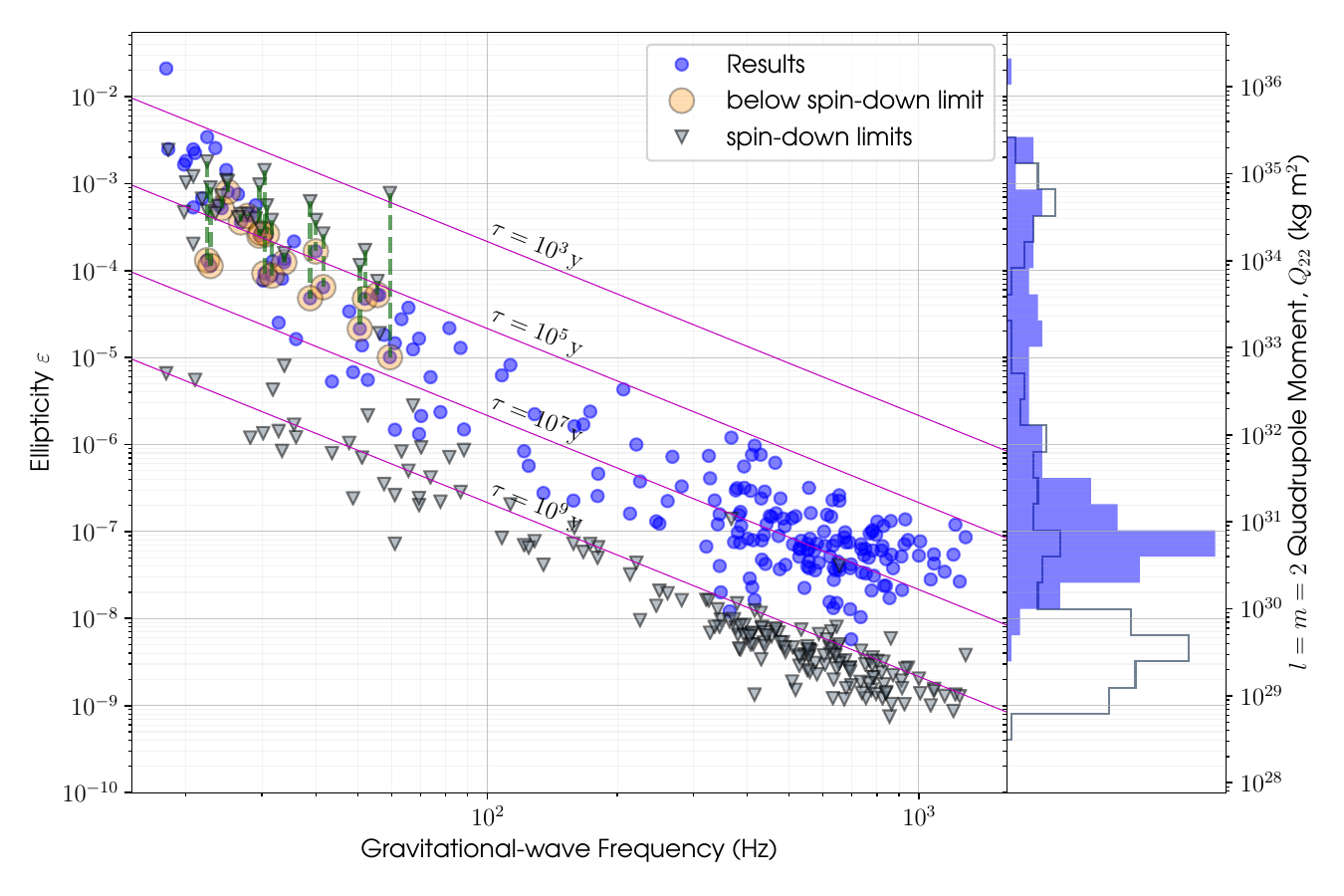}
\caption{Upper limits on mass quadrupole $Q_{22}$ and fiducial ellipticity $\varepsilon$ for
221 pulsars. The filled circles show the limits as derived from the observed upper limits on
the gravitational-wave amplitude $h_0$ assuming the canonical moment of inertia and distances. Triangles show the limits based derived from each
pulsar's observed spin-down. The diagonal lines show contours of equal characteristic age $\tau$
assuming that braking is entirely through gravitational-wave emission. The distributions of these limits are also
show in histogram form to the right of the figure, with the filled and unfilled histograms showing
our observed limits and the spin-down limits, respectively.\label{fig:ell}}
\end{figure*}

\end{document}